\documentclass[aps,prb,twocolumn,groupedaddress,showpacs]{revtex4}

\usepackage{graphicx}

\def\C60{C$_{60}$}
\def\A4C60{A$_4$C$_{60}$}
\def\K4C60{K$_4$C$_{60}$}
\def\Rb4C60{Rb$_4$C$_{60}$}
\def\Cs4C60{Cs$_4$C$_{60}$}
\def\cm-1{cm$^{-1}$}
\def\cn{C$_{60}^{4-}$}
\def\T1u{$T_{1u}$}

\def\Dhd{$D_{3d}$}
\def\Dod{$D_{5d}$}
\def\D2h{$D_{2h}$}
\def\deg{$^{\circ}$}

\begin{document}

\title{Static and dynamic Jahn-Teller effect in the alkali metal fulleride salts A$_4$C$_{60}$\\ (A = K, Rb, Cs)}

\author{G. Klupp}
\email[]{klupp@szfki.hu}
\affiliation{Research Institute for Solid State Physics and
Optics, Hungarian Academy of Sciences, P. O. Box 49, H-1525
Budapest, Hungary}
\author{K. Kamar\'as}
\email[]{kamaras@szfki.hu}
\affiliation{Research Institute for Solid State Physics and
Optics, Hungarian Academy of Sciences, P. O. Box 49, H-1525
Budapest, Hungary}
\author{N. M. Nemes}{
\altaffiliation{Present address: Instituto de Ciencia de Materiales de Madrid (ICMM-CSIC),
28049 Cantoblanco, Madrid, Spain}
  \affiliation{NIST Center for Neutron Research, Gaithersburg, MD 20899-8562, USA}
  \affiliation{Department of Materials Science and Engineering,
  University of Maryland, College Park, MD 20742, USA}

\author{C. M. Brown}{
  \affiliation{NIST Center for Neutron Research, Gaithersburg, MD 20899-8562, USA}
  \affiliation{Department of Materials Science and Engineering,
  University of Maryland, College Park, MD 20742, USA}
\author{J. Le\~{a}o }{
  \affiliation{NIST Center for Neutron Research, Gaithersburg, MD 20899-8562, USA}

\date{\today}
\begin{abstract}

We report the temperature dependent mid- and near-infrared spectra
of \K4C60, \Rb4C60 and \Cs4C60. The splitting of the vibrational
and electronic transitions indicates a molecular symmetry change
of \cn\ which brings the fulleride anion from \D2h to either a
\Dhd\ or a \Dod\ distortion. In contrast to \Cs4C60, low
temperature neutron diffraction measurements did not reveal a
structural phase transition in either \K4C60 and \Rb4C60. This
proves that the molecular transition is driven by the molecular
Jahn-Teller effect, which overrides the distorting potential field
of the surrounding cations at high temperature. In \K4C60 and
\Rb4C60 we suggest a transition from a static to a dynamic
Jahn-Teller state without changing the average structure. We
studied the librations of these two fullerides by temperature
dependent inelastic neutron scattering and conclude that both
pseudorotation and jump reorientation are present in the dynamic
Jahn-Teller state.

\end{abstract}


\maketitle

\section{\label{intro}Introduction}

The insulating character of the \A4C60 (A  = K, Rb, Cs) compounds
has been a longstanding puzzle in fullerene science. The
successful description involves a combination of the molecular
Jahn-Teller (JT) effect and the Mott-Hubbard band picture
resulting in the theory of the nonmagnetic Mott--Jahn--Teller
insulating state.\cite{fabrizio97} This theory has been used
effectively for the explanation of EELS\cite{knupfer96,knupfer97}
and NMR\cite{brouet02} measurements on \A4C60. Recently, a
sophisticated experiment\cite{wachowiak05}  by scanning tunneling
microscopy has revealed JT distorted molecules in \K4C60\ monolayers.
In macroscopic crystals, however, the
distortion could only be detected directly in one case: anions
with \D2h symmetry were found in \Cs4C60 by neutron
diffraction.\cite{dahlke02} In \Cs4C60, X-ray\cite{dahlke98} and
neutron diffraction measurements also found an
orthorhombic-tetragonal ($Immm$ to $I4/mmm$) phase transition
between 300 and 623 K. The crystal structure of \K4C60\ and
\Rb4C60\ was determined to be $I4/mmm$ at room
temperature,\cite{kuntscher97,bendele98} although atomic positions
were not refined. In the case of \K4C60 mid-infrared (MIR) and
near-infrared (NIR) measurements showed a splitting that indicated
a JT distorted anion.\cite{iwasa95} The distortion was found to be
temperature dependent\cite{kamaras02} and the possibility of a
similar phase transition as that in \Cs4C60 has been put forward.

Vibrational spectroscopy is uniquely sensitive to the change in
molecular symmetry (\emph{i.e.} the exact shape of the molecule) through
the splitting of vibrational bands.  Because it detects the motion
of atoms, it naturally goes beyond the spherical approximation
used for crude models of the electronic structure. In this
respect, molecular vibrations are more intimately connected to
structural studies which show the average position of the atomic
cores, than to methods probing magnetic and electronic excitations
where an analogy to atomic orbitals is often sufficient to
describe the results. In this paper, we follow the distortions of
fulleride ions in three \A4C60 salts (A=K, Rb, Cs) with
temperature. Our conclusions are mainly drawn from mid-infrared
vibrational spectra, but we also study the effect of these
distortions on electronic orbitals of the \cn\ ions, through NIR
spectra probing both intra- and intermolecular electronic
excitations. To clarify whether the distortions are caused by
crystal potential or molecular degrees of freedom, we performed
temperature-dependent neutron diffraction studies, complemented by
inelastic neutron scattering in order to detect molecular motion.
We find no structural phase transition to a cooperative static
Jahn-Teller state in either \K4C60 or \Rb4C60 down to 4K; changes
in vibrational spectra reflect the change in molecular symmetry
and thus a transition from static to dynamic Jahn-Teller state as
the temperature is raised.

\section{The Jahn-Teller effect in fulleride salts}

To understand the precise nature of the distortions occurring in
fulleride salts, we have to elaborate on the details of their
crystal and molecular symmetry. An entire monograph has been
devoted to this question,\cite{konyv} here we will only repeat the
main statements.

The Jahn-Teller effect is caused by the interaction of a
degenerate electronic state with molecular
vibrations.\cite{jahnteller37} In \C60 anions, the electronic
states involved are those of the triply degenerate $t_{1u}$
orbitals and the vibrations are the ten fivefold degenerate $H_g$
modes. These interactions result in a change of shape of the
molecule and consequently a change in the splitting of the
electronic orbitals. Electrons will occupy the lowest-energy
levels and thus, if the splitting is large enough, overcome Hund's
rule and form nonmagnetic systems.

In most solids containing open-shell species, the energy bands
are much broader than the JT splitting; this is why
the A$_3$C$_{60}$
salts are metals.\cite{hebard91} In this case the electrons cannot be
assigned to individual molecules and therefore Jahn-Teller
coupling is not possible. The insulating character of \A4C60\
salts has been proposed to be caused by Mott localization which
enables Jahn-Teller coupling between the localized electrons and
the vibrational degrees of freedom. This state has been termed the
"Mott-Jahn-Teller nonmagnetic insulator".\cite{fabrizio97} As we
will see, our results fully support this picture, so we describe
the \A4C60\ systems in this framework.

In the atomic orbital-like classification used by Chancey and
O'Brien,\cite{konyv} the \cn\ molecular ion is a $p^4 \otimes h$
system, where the allowed Jahn--Teller distortions for isolated
ions are \D2h, \Dhd\ and \Dod. The predicted shape of the
distortions is "pancake-type": a flattening along a $C_2$, $C_3$
or $C_5$ molecular axis.\cite{konyv} We illustrate these possible
distortions in Figure \ref{fig:distort}a. The adiabatic potential
energy surface (APES) of these systems has minima at either \Dhd\
or \Dod\ symmetry, and saddle points at \D2h\
symmetry.\cite{obrien96}  If the \Dhd\ geometries are minima, then
the \Dod\ are maxima and vice versa. There are six possible \Dod\
distortions and ten \Dhd\ distortions in different directions;
transitions between them occur through tunneling which results in
a different molecular shape without the rotation of the molecule
itself.\cite{dunn95} This motion is called \emph{pseudorotation}.
Proof of such dynamic distortions has indeed been presented
recently by a sophisticated experiment on monoanions produced in a
storage ring.\cite{tomita05} The \D2h distortion can only be
realized when an external potential lowers the energy of this
distortion. Forming a solid from fulleride ions with counterions
creates such a potential field.

\begin{figure}
\includegraphics[scale=0.35]{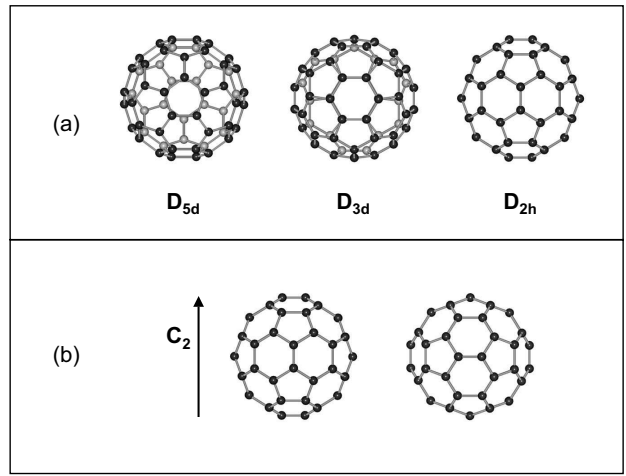}
\caption{(a) Possible Jahn-Teller distortions in fulleride ions.
The axis along which the distortion occurs is perpendicular to the
plane of the paper through the center of the molecule. The
atoms above the plane of the paper are black while those under it are
grey. (b) The two
standard orientations in fullerene/fulleride solids. The \emph{z}
crystal axis coincides with the $C_2$ molecular axis shown. } \label{fig:distort}
\end{figure}

Apart from the intrinsic JT distortion, forcing an icosahedral
\C60\ molecule into a crystal inevitably
lowers its symmetry. For all crystal systems with orthogonal
principal axes and one molecule per primitive unit
cell the $C_2$ symmetry axes of the molecule are aligned with the
crystallographic axes, but the molecule can assume two different
orientations as shown in Figure \ref{fig:distort}b. These are the standard orientations
originally defined for
orientationally ordered \C60.\cite{harris92} Thus, for a C$_{60}^{n-}$ anion the $x$ and $y$
molecular axes are not equivalent, reflecting the lack of a
fourfold axis in icosahedral symmetry. Nevertheless, the
misconception prevailed in the early fullerene literature that in
a tetragonal distorting field, the \emph{individual molecular
ions} can be uniaxially distorted into the $D_{4h}$
pointgroup with the $c$ crystal axis as principal
axis.\cite{lukyanchuk95,lacilaci01,kerkoud96} This approach treats the fulleride ions as a
sphere\cite{ozaki86} (a "giant atom"), and takes the effect of the
distorting crystal field to be the same as the inherent JT
distortion of the balls, leading to the conclusion that the two
are indistinguishable. It is apparent from structural studies,
however,\cite{kuntscher97,bendele98} that in a tetragonal system
the \C60\ molecules cannot be equivalent. Orbital overlap between
cations and anions determines whether the balls are ordered or
disordered, \emph{i.e.} the crystal is tetragonal or
orthorhombic,\cite{fischer93,dahlke02,yildirim93b} but
the molecular symmetry is the same \D2h in both
cases. The consequence is that
the threefold degenerate orbitals will show a threefold splitting
in both an orthorhombic and tetragonal environment. An
orthorhombic crystal is formed by simply arranging the D$_{2h}$
distorted molecules in an ordered fashion, while the overall
tetragonal symmetry of the crystal can only be maintained as an
average with some sort of disorder, either static or
dynamic.\cite{bendele98} In the following we try to summarize the
possible arrangements and relate them to the crystal structures as
classified by Fabrizio and Tosatti.\cite{fabrizio97} Note that we
consciously avoid the use of the term \emph{merohedral
disorder} throughout the discussion: this concept is correctly used
for A$_3$C$_{60}$ systems\cite{stephens91} but is incorrect for \A4C60.

\begin{figure*}
\includegraphics[scale=0.5,angle=-90,trim=40 0 0 0]{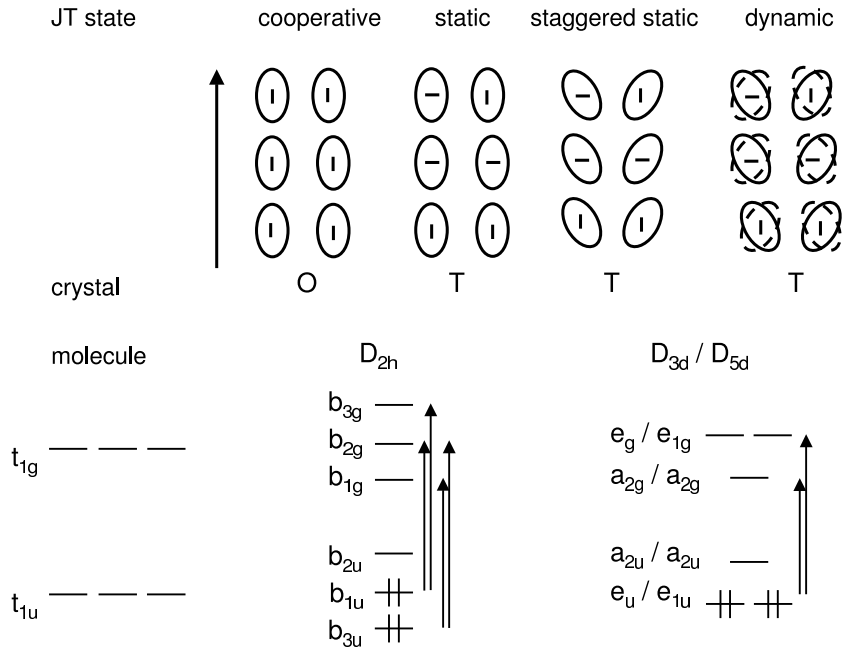}
\caption{Top: Some possible arrangements of fulleride ions in the
crystal structures in \A4C60 systems and their molecular
distortions. Ellipses with horizontal and vertical bars,
respectively, represent the standard orientations in Figure
\ref{fig:distort}b. O stands for orthorthombic, T for tetragonal
crystal structure. Bottom: splitting of the HOMO ($t_{1u}$) and
the LUMO  ($t_{1g}$) molecular orbitals of \cn\ in a \D2h\
(cooperative static, static)  and \Dhd\ or \Dod\ (staggered
static, dynamic) distorted molecule, respectively. Infrared-active
vibrations of $T_{1u}$ symmetry show similar splitting. Arrows
denote dipole allowed transitions between split
states.}\label{fig:JTstate}
\end{figure*}

The ordered orthorhombic structure mentioned above is the
so-called \emph{cooperative static Jahn-Teller state}. For the overall
symmetry to become tetragonal, we have to assume there exists an
average (spatial or temporal) of several molecules. We summarize
the situation in Figure \ref{fig:JTstate}. The static disorder
means that all molecules align their $C_2$ axes in the \emph{c}
direction, but the hexagon-hexagon bonds are
randomly oriented along either the \emph{a} or \emph{b} axes
(\emph{static Jahn-Teller state}). This scenario is, however, not
the only possible geometry whose spatial average results in a
tetragonal crystal. Molecules distorted either along the $C_3$ or
the $C_5$ axis can form  an ordered array resulting in a fourfold
axis in the $c$ direction (\emph{staggered static Jahn-Teller
state}). The molecular principal axis of a \Dhd\ or \Dod\ anion
cannot be parallel with the crystallographic axes, but must be
arranged such that the overall average gives an $I4/mmm$
structure. The transition from static to staggered static state
occurs through pseudorotation and vibration, \emph{i.e.} the coordinates
of the individual carbon atoms change only slightly and there is
no reorientation of the molecule as a whole. If there
are several configurations with small energy barriers between them
(compared to the energy of thermal motion), the balls can assume
many of these configurations dynamically and thus the
\emph{dynamic Jahn-Teller state} is formed. The significance of
pseudorotation increases as the amplitude of thermal motion
becomes larger.

\begin{figure*}
\includegraphics[scale=0.5]{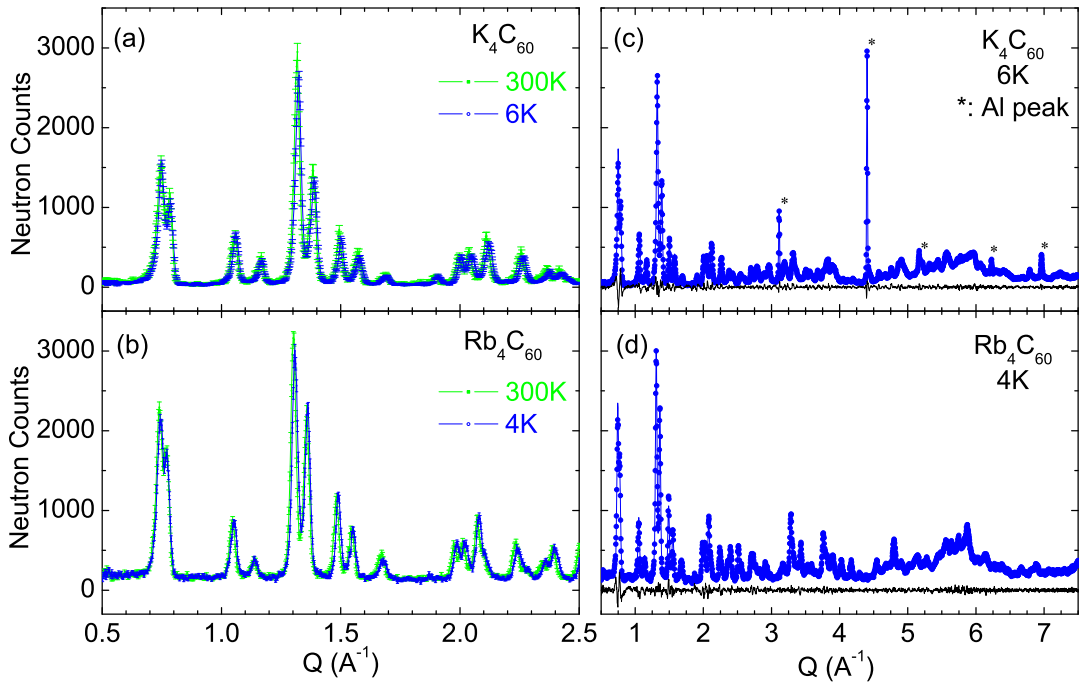}
\caption{(color online) Neutron diffraction profiles of (a)
\K4C60\ and (b) \Rb4C60\ at two temperatures, showing only a small
thermal contraction of the lattice but no change in symmetry or
unit cell. LeBail fits for lattice parameter determination for (c)
\K4C60 at 6~K and (d) \Rb4C60 at 4~K. The measured data on these
latter two graphs are shown by blue dots, the fits by blue lines
and the difference plots by black lines.} \label{fig:ndiffr}
\end{figure*}

The detection of the distortion by diffraction methods demands
extreme sensitivity, as the magnitudes in question are small. Paul
\emph{et al.}\cite{paul94} found a quasi-axial elongation of 0.04 {\AA} in
(PPN)$_2$C$_{60}$, where the symmetry of the C$_{60}^{2-}$ dianion
is lowered to $C_2$. In the monovalent decamethylnickelocenium
salt,\cite{wan95} the symmetry was found close to \D2h, with a
difference between maximum and minimum radii of 0.05 {\AA}. These
are static distortions in which the role of the bulky organic
counterions and the inherent JT effect cannot be
separated.\cite{reed00} The largest distortion (defined as the
difference between the smallest and the largest distance from the
center of the ball) so far has been found in the ordered
orthorhombic phase of \Cs4C60; 0.076 \AA\ at 300 K.\cite{dahlke02}
In \K4C60 Kuntscher et al.\cite{kuntscher97} put
an upper limit of 0.04 \AA\ on the difference between "equatorial"
and "polar" radii. One must take into account, though,
that in the case of a staggered static arrangement the directions of
maximum and minimum radii are not necessarily the crystal axes,
and in the dynamic case the difference is smeared out completely.

In Figure \ref{fig:JTstate}(bottom), we show the corresponding
splitting of the molecular orbitals. In the icosahedral \C60\
molecule the LUMO (which becomes the HOMO in the molecular ions)
is a threefold degenerate $t_{1u}$ orbital. We obtain a threefold
splitting of this orbital in the cooperative static and static JT states
and a twofold splitting in the staggered static and dynamic
states. The lowest-energy final states for
dipole transitions also derive from a threefold degenerate
orbital, the even-parity $t_{1g}$ one. Incidentally, the four
infrared-active vibrations of \C60\ also show $T_{1u}$ symmetry,
therefore the discussion can proceed along the same lines.

Further complications arise if we take into account that the
fullerene balls are capable of rotation around several axes. The
simplest scenario would be that occasional reorientational jumps
between the two standard orientations (around the $C_2$ axis)
would average out the symmetry from \D2h\ to $D_{4h}$. We know
from inelastic neutron scattering in
K$_3$C$_{60}$,\cite{neumann93} however, that the rotation between
standard orientations occurs with a much higher probability around
a $C_3$ axis. Structural studies and modeling\cite{kuntscher97} in
\A4C60 indicated that rotation around the $C_2$ axis is hindered
because of unfavorable alkali atom -- carbon distances. Thus, we
suggest that dynamic disorder in \A4C60 salts is the result of
reorientation around the threefold axes.

\section{Experimental}

\A4C60 systems have been prepared previously either by solid-state
synthesis\cite{fleming91, murphy92, poirer95} or by a liquid
ammonia route.\cite{dahlke98} We used a solid-state synthesis for
all three alkali salts by reacting stoichiometric amounts of the
alkali metal with \C60 at 350 $^{\circ}$C in a steel capsule. The
reaction was followed using powder X-ray diffraction and MIR
spectroscopy. The reaction mixture required heating for 10 to 14
days with one intermediate sample regrinding in the case of \K4C60
and \Cs4C60, and 20 days with three regrindings for \Rb4C60, to
achieve complete conversion. No impurities were observed in the
\K4C60 and \Rb4C60 samples, while X-ray diffraction found less
than 5 \% Cs$_6$C$_{60}$ in \Cs4C60.

Since fullerides are air sensitive, the synthesis and sample
preparation was conducted in a dry box. For the MIR and NIR
measurements, KBr pellets were pressed and transmittance spectra measured
with the sample inside a liquid nitrogen cooled flow-through
cryostat under dynamic vacuum. Spectra were recorded with resolution of 1 or 2 \cm-1
in the MIR range using a Bruker IFS 28 spectrometer and 4
\cm-1 in the NIR using a Bruker IFS 66v/S spectrometer.

Neutron scattering measurements were performed at the NIST Center
for Neutron Research. Large amounts of materials were prepared for
these experiments (2.4~g of \K4C60\ and 1.1~g of \Rb4C60) to
achieve good counting statistics. Temperature dependent
neutron diffraction data were collected on the BT1 diffractometer
using a wavelength of $\lambda$ = 1.5403 {\AA} and a Q-range of
0.2 {\AA}$^{-1}$--8.1 {\AA}$^{-1}$ with the Cu(311) monochromator
set at a 90° take-off angle and  using in-pile collimation of 15
minutes of arc. Lattice parameters were extracted using the LeBail
method.\cite{GSAS,EXPGUI,LeBail}

Low energy molecular librations were studied using the BT4 triple-axis
spectrometer. We collected constant momentum transfer(Q) scans at Q=5.5 {\AA}$^{-1}$
with a fixed incident energy of 28~meV. The incident
beam was produced using a Cu(220) monochromator and a graphite
filter for removal of higher order contamination.
The scattered beam was analyzed using a graphite(004)
crystal. The measured resolution with 60'-40'-40'-40'
collimation was 0.97~meV full width at half maximum. Samples
were loaded in indium-wire-sealed aluminum and vanadium
cylindrical cans. Sample temperature was controlled between 4~K and 300~K
with a closed cycle helium refrigerator. In the analysis of librational
spectra, background runs were first subtracted, the intensities
were corrected for changes in the scattered energy contribution to
the spectrometer resolution, and then the spectra were symmetrized
for the thermal Bose factor. The corrected data were subsequently
fitted using a Gaussian resolution function at zero energy
transfer and two identical Lorentzians symmetrically located about the elastic line,
and convoluted with the instrumental resolution function.
Details of similar librational studies on
other fullerides can be found in Ref.~\onlinecite{neumann93}.

\section{Results}

\subsection{Structure}

Room temperature X-ray diffraction showed the crystal symmetries
to be $I4/mmm$ in \K4C60\ and \Rb4C60, and $Immm$ in \Cs4C60, in agreement with published
results.\cite{kuntscher97,dahlke98} Temperature-dependent
structural studies were reported only for \Cs4C60,\cite{dahlke02}
revealing an orthorhombic-tetragonal transition between 300 K and
623 K. Previously, based on vibrational spectra in
\K4C60,\cite{kamaras02} we suggested a similar transition in the
two other alkali compounds. In order to draw  a definitive
conclusion on this hypothesis, we performed low-temperature
neutron diffraction measurements on \K4C60\ and \Rb4C60. The
resulting low- and high-temperature diffraction patterns are
compared in Figure \ref{fig:ndiffr}. We found that the structure of both \K4C60\ and
\Rb4C60\ remain tetragonal down to the lowest temperatures
measured. The 4 K and 300 K lattice parameter values from the LeBail
analysis are given explicitly in
Table~\ref{table:lattice-const}. The room-temperature data agree with those
of Kuntscher \emph{et al}.\cite{kuntscher97} for \K4C60\ and
\Rb4C60.

We investigated the \K4C60\ salt in detail, at several
temperatures, to make sure we did not miss a possible
tetragonal-tetragonal phase transition, similar to the one found
in \Rb4C60 with increasing pressure.\cite{sd04} The lattice
parameters extracted at each temperature and normalized to the
room-temperature values are shown in Figure
\ref{fig:lattice-const}. As no significant change can be seen
except for a small thermal contraction (an order of magnitude smaller
than the pressure-induced change reported in Ref. \onlinecite{sd04}), we can rule out even a
tetragonal-tetragonal phase transition in \K4C60.

\begin{figure}
\includegraphics[scale=0.31,angle=-90,trim=0 0 0 0]{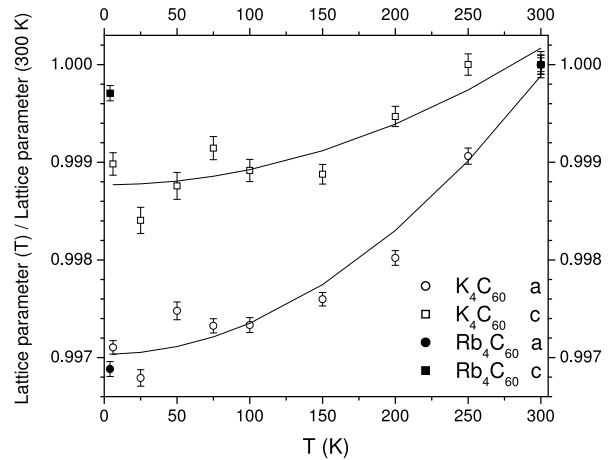}
\caption{Temperature-dependent lattice parameters normalized to
the 300~K values of \K4C60 and \Rb4C60. The lowest and highest
temperature values are given in Table~\ref{table:lattice-const}.
The solid lines are guides to the eye.} \label{fig:lattice-const}
\end{figure}

\begin{table}
\caption{Lattice parameters of \K4C60 and \Rb4C60\ at the lowest and
highest measured temperatures.} \label{table:lattice-const}
\begin{ruledtabular}
\begin{tabular}{l|ll|ll}
T&\multicolumn{2}{c|}{\Rb4C60}&\multicolumn{2}{c}{\K4C60}\\
    &a (\AA) &c (\AA) & a (\AA) & c (\AA) \\
\hline 4-6~K & 11.912(1) & 11.007(1) & 11.827(1) & 10.746(1) \\
      300~K & 11.949(1) & 11.011(1) & 11.862(1) & 10.757(1) \\
\end{tabular}
\end{ruledtabular}
\end{table}

%\begin{table}
%\caption{Lattice parameters of \K4C60 and \Rb4C60\ at the lowest
%and highest measured temperatures.}
%\label{table:lattice-const}
%\begin{ruledtabular}
%\begin{tabular}{l|ll|ll}
%T(K)&\multicolumn{2}{c|}{\Rb4C60}&\multicolumn{2}{c}{\K4C60}\\
%     &a&c&a&c \\
%\hline 4-6~K & 11.912(1) & 11.007(1) & 11.827(1) & 10.746(1) \\
%       300~K & 11.949(1) & 11.011(1) & 11.862(1) & 10.757(1) \\
%\end{tabular}
%\end{ruledtabular}
%\end{table}

\subsection{Molecular vibrations}

\begin{figure}
\includegraphics[scale=0.305,trim=0 0 0 0]{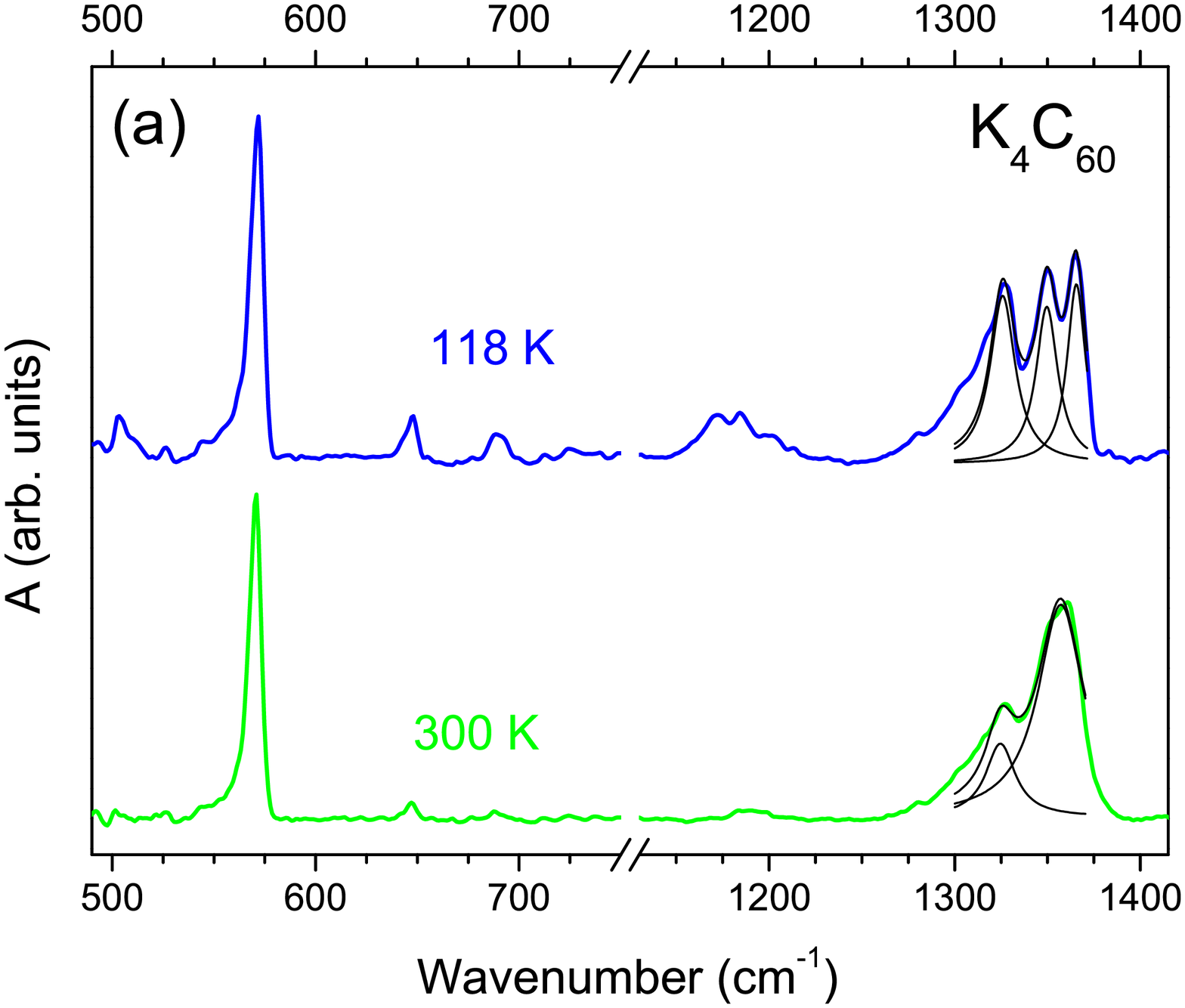}
\includegraphics[scale=0.32]{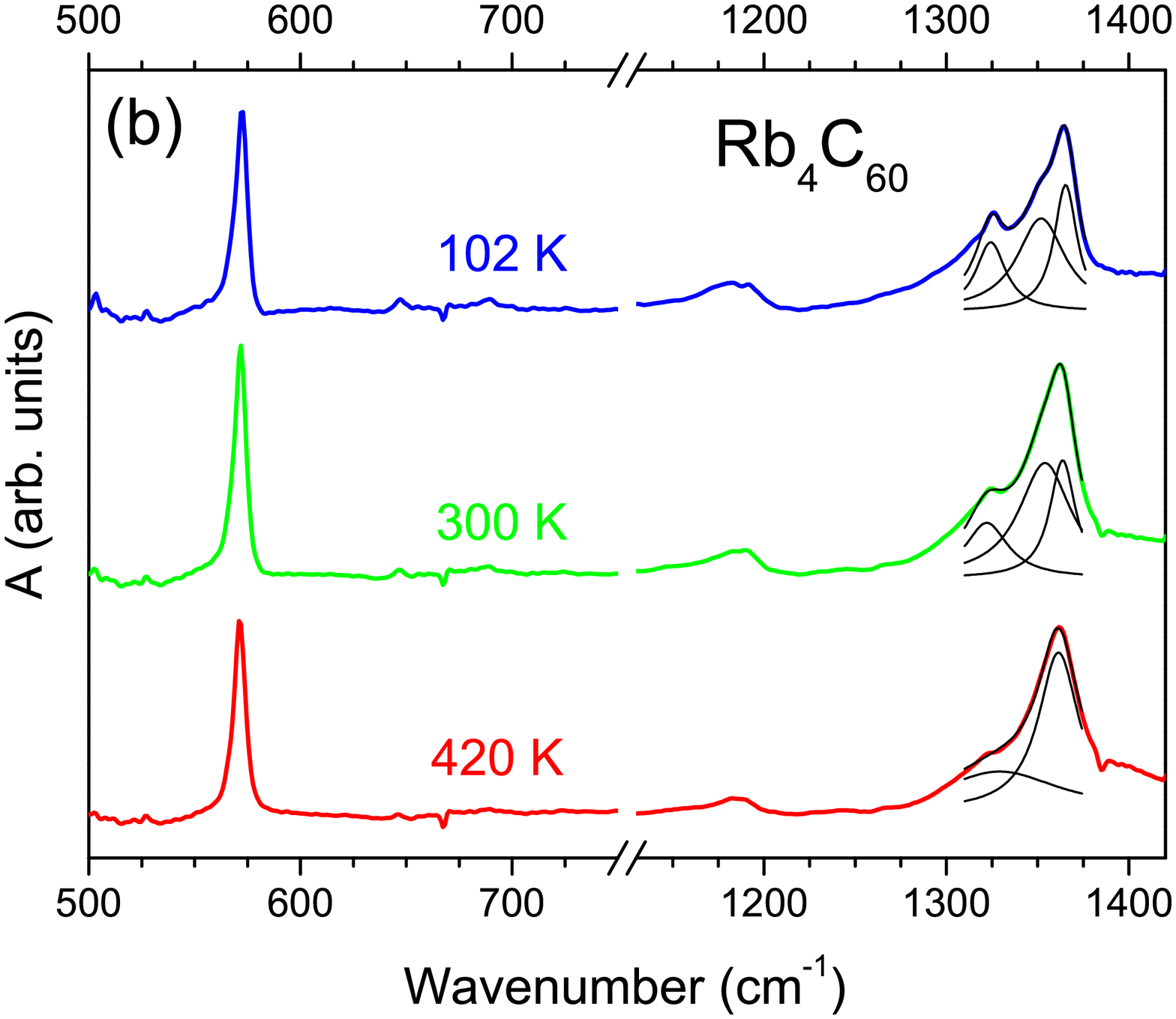}
\includegraphics[scale=0.32]{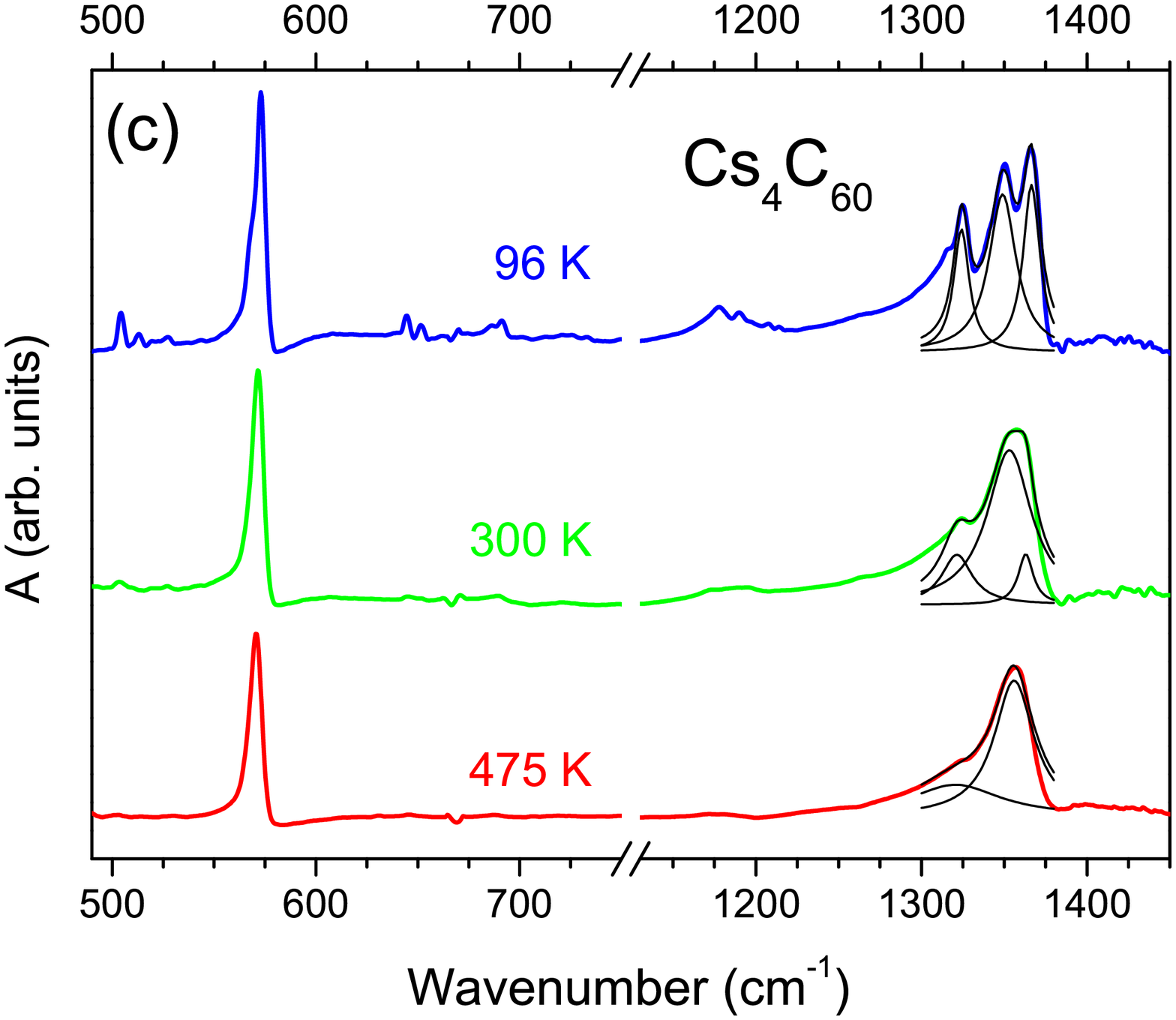}
\caption{(color online) MIR spectrum of (a) \K4C60, (b) \Rb4C60,
and (c) \Cs4C60\ at selected temperatures above and below the
change in symmetry (colored lines).  The highest-frequency mode
can be fitted with three Lorentzians at low temperature and two
Lorentzians at high temperature (black lines). These splittings
indicate a molecular symmetry change with
temperature.}\label{fig:mir}
\end{figure}

The MIR spectra of the three salts measured at room temperature
and at characteristic temperatures unique for each salt are shown
in Figure \ref{fig:mir}. Since \C60 is an icosahedral molecule, it
has only four infrared active vibrations (at 528, 577, 1183 and
1429 \cm-1)\cite{kratschmer90}, all of which belong to the \T1u
representation. The shift and the splitting of the highest
frequency \T1u(4) mode has been used as the most sensitive
indicator for charge transfer,\cite{pichler94} symmetry
change\cite{kamaras97} and bonding\cite{rao97} in fullerene
compounds. The most prominent feature of our spectra is the
splitting of this mode (shifted to 1350~\cm-1 because of charge
effects) indicating a lowering of symmetry from icosahedral. All
spectra could be fitted with either two or three Lorentzians in
this frequency range, and the results are summarized in Table
\ref{tab:mirfit}. The temperature dependence of the splitting of
the \T1u(3) mode around 1182~\cm-1 was found to be the same as for
the \T1u(4) mode. In contrast, the two lower-frequency modes were
not split at our resolution, instead, we observed a decrease in
peak height and increase in linewidth of the \T1u(2) mode at
571~\cm-1 (Fig. \ref{fig:irhom}). (We note that in pristine \C60
below the orientational phase transition\cite{homes94} no
splitting was observed in the \T1u(2) mode even at 0.4 \cm-1
resolution and the splitting of the \T1u(1) mode was less than 1
\cm-1. The latter mode is almost unobservable in the \cn\
ion.\cite{pichler94}) The symmetry lowering from $I_h$ also
activates previously silent modes, appearing between 600-750
\cm-1. The intensity of these peaks increases on cooling. The
temperature dependence of this increase can also be used to follow
the symmetry change of the \cn anion.

\begin{table*}
\caption{The parameters of the Lorentzians fitted to the \T1u(4)
mode (wavenumber: $\nu^*$, full width at half maximum: $w$,
integrated intensity: $I$). The intensities were normalized to the
sum of intensities at 300~K. \label{tab:mirfit}}
\begin{ruledtabular}
\begin{tabular}{l|ll|lll|lll}
 & \multicolumn{2}{c|}{\K4C60} &
\multicolumn{3}{c|}{\Rb4C60}& \multicolumn{3}{c}{\Cs4C60}
\\  & 118~K& 300~K& 102~K& 300~K& 420~K&96~K & 300~K & 475~K
\\ \hline $\nu^*_1$(\cm-1) & 1326 $\pm$ 1 & 1324 $\pm$ 1& 1324 $\pm$ 1& 1322 $\pm$ 1& 1330 $\pm$
7&1324 $\pm$ 1& 1321 $\pm$ 1& 1323 $\pm$ 3
\\ $w_1$(\cm-1) & 15 $\pm$ 2& 17 $\pm$ 4& 19 $\pm$ 2& 28 $\pm$ 3& 82 $\pm$
31&11 $\pm$ 2 & 21 $\pm$ 2& 55 $\pm$ 12
\\ $I_1$ & 3 $\pm$ 1 & 2 $\pm$ 1 & 2 $\pm$ 1 & 2 $\pm$ 1 & 4.3 $\pm$
4.1&4 $\pm$1 & 3$\pm$1 & 4$\pm$3
\\ \hline $\nu^*_2$(\cm-1) & 1350 $\pm$ 1& 1358 $\pm$ 1& 1352 $\pm$ 1& 1354 $\pm$ 2& 1361 $\pm$
1&1349 $\pm$ 1& 1354 $\pm$ 1& 1356 $\pm$ 1
\\ $w_2$(\cm-1) & 14 $\pm$ 2& 33 $\pm$ 2& 32 $\pm$ 4& 33 $\pm$ 3& 27 $\pm$
2&20 $\pm$ 2 & 34 $\pm$ 2& 27 $\pm$ 2
\\ $I_2$ & 3$\pm$1 & 9$\pm$3 & 2$\pm$1 & 5$\pm$2 & 5$\pm$2&9$\pm$2 & 15$\pm$2 & 10$\pm$1
\\ \hline $\nu^*_3$(\cm-1) & 1365 $\pm$ 1&  & 1365 $\pm$ 1& 1363 $\pm$
1& &1366 $\pm$ 1& 1363 $\pm$ 1&
\\ $w_3$(\cm-1)& 12 $\pm$ 2& & 15 $\pm$ 2& 17 $\pm$ 2& &12 $\pm$ 2& 10 $\pm$ 2&
\\ $I_3$ & 3$\pm$1 &  & 2$\pm$1 & 3$\pm$1 & &6$\pm$1 & 1.2$\pm$0.7 &
\\
\end{tabular}
\end{ruledtabular}
\end{table*}

From Figure \ref{fig:JTstate}, it follows that the \T1u modes
split twofold in the \Dod\ and \Dhd\ point groups, and threefold
in the \D2h\ point group and since all split modes are IR active,
this directly indicates the  geometry. It is also clear that in
all compounds, the distortion changes from the latter to the
former upon warming. The temperature where this occurs depends on
the counterion. From Figure \ref{fig:irhom} the transition
temperatures for \K4C60, \Rb4C60\ and \Cs4C60\ are approximately
270 K, 330 K and 400 K, respectively, increasing with increasing
cation size.

\begin{figure*}

\includegraphics[scale=0.25,angle=0,trim=75 250 75 -300]{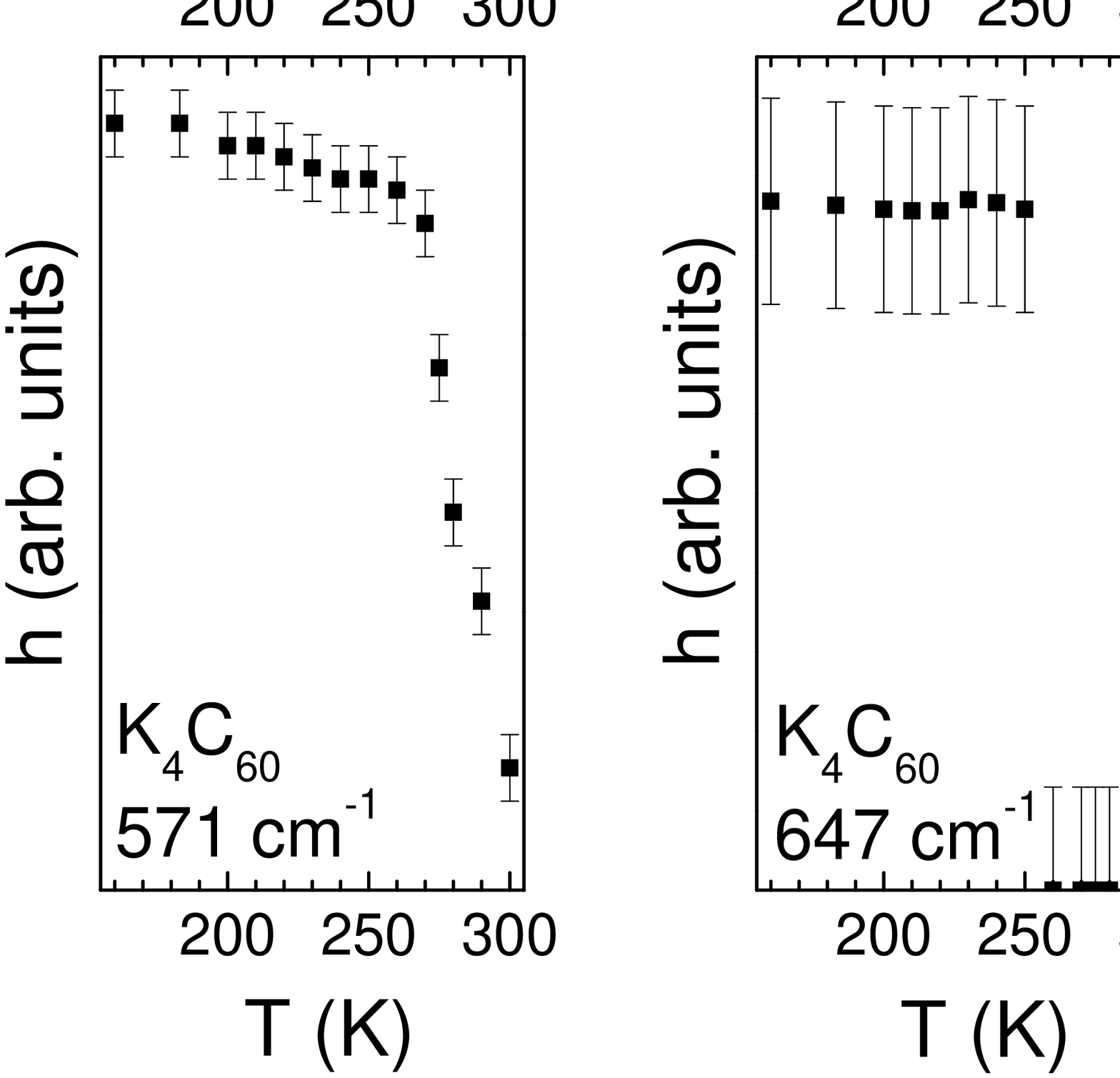}
\includegraphics[scale=0.25,angle=0,trim=75 250 100 -300]{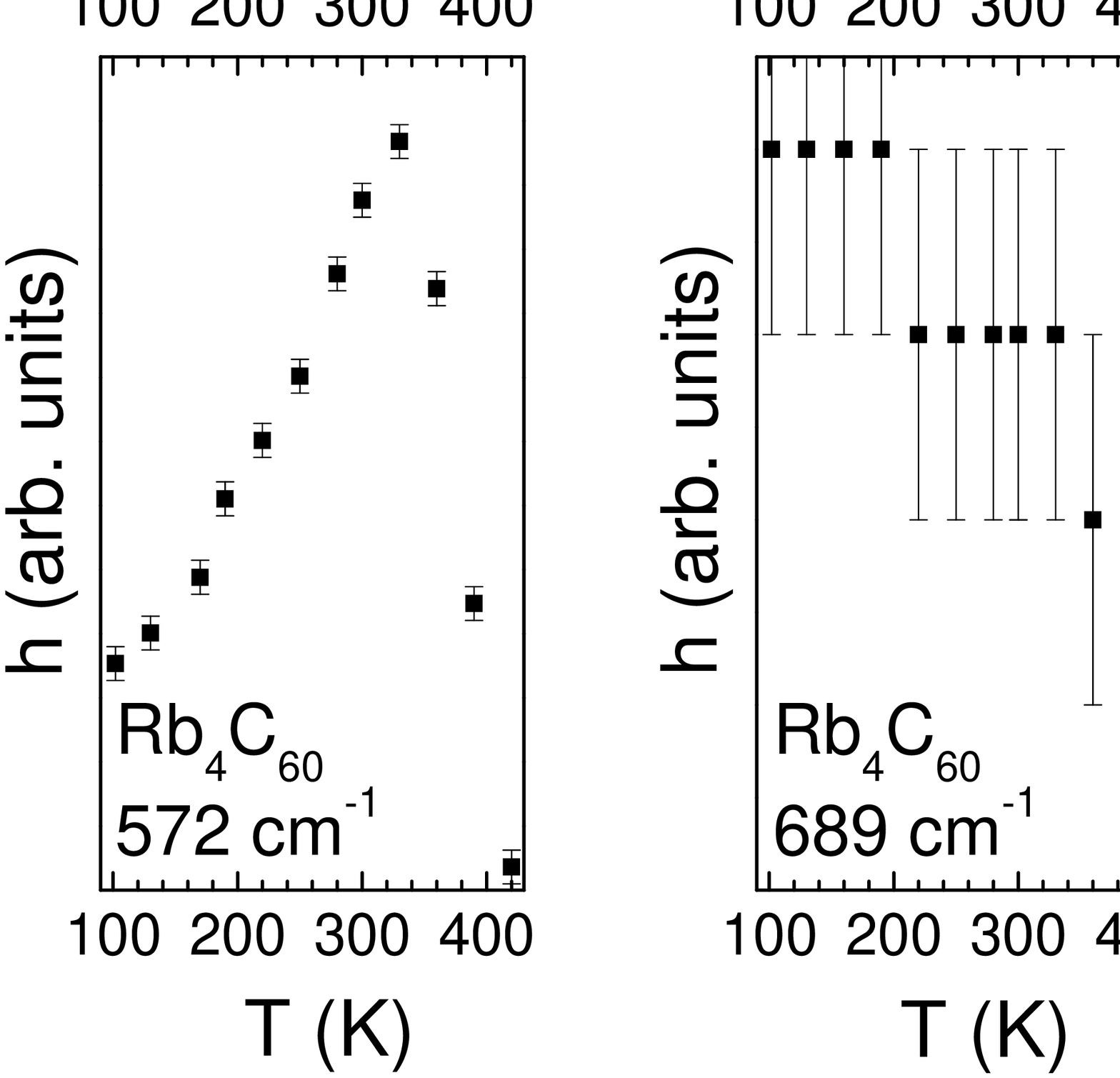}
\includegraphics[scale=0.25,angle=0,trim=75 250 125 -300]{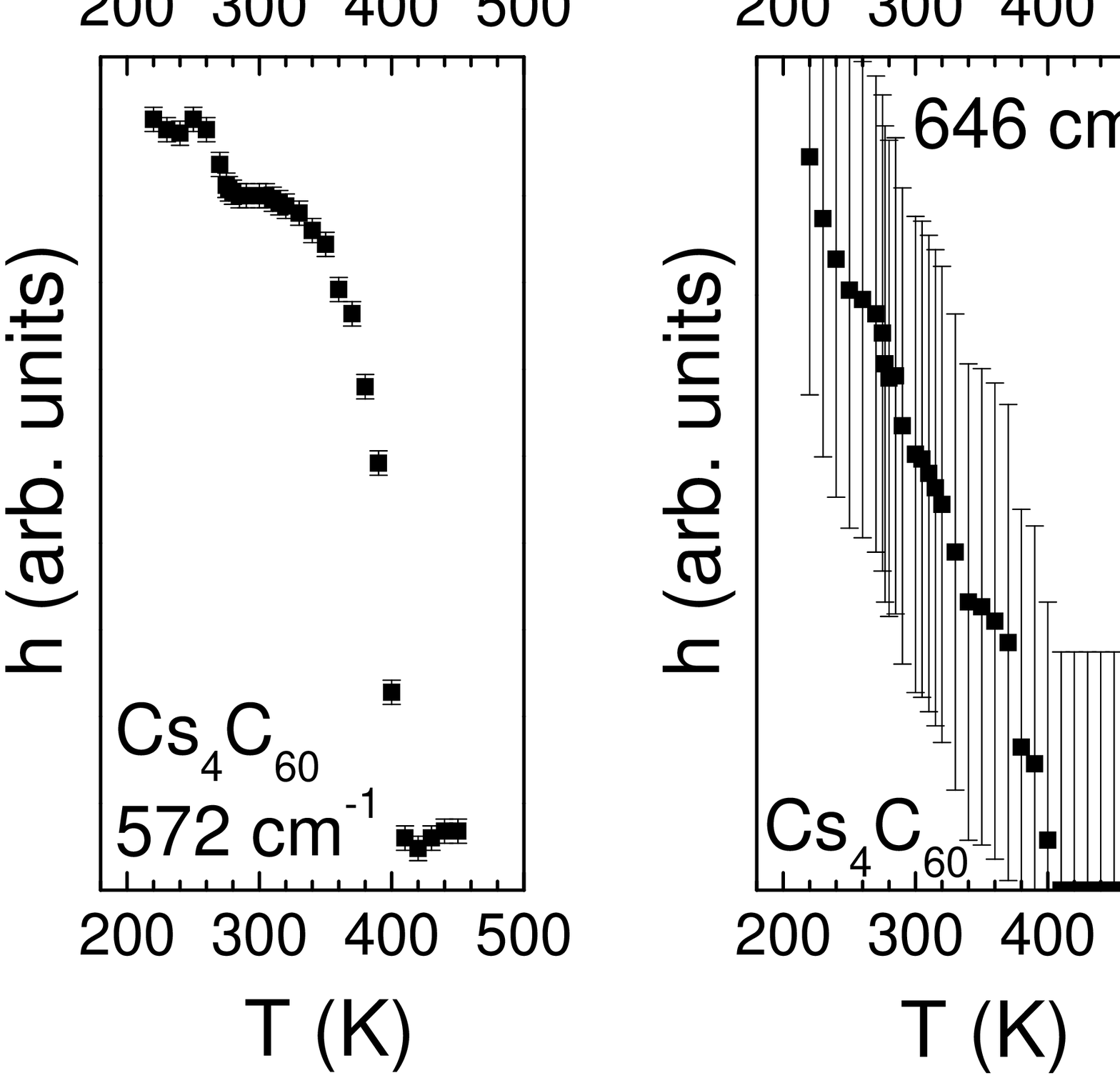}
\caption {Temperature dependence of the peak heights of selected
MIR lines in \K4C60, \Rb4C60, and \Cs4C60. Changes are apparent
around 270, 330 and 400~K, respectively.} \label{fig:irhom}
\end{figure*}

\begin{figure}
\includegraphics{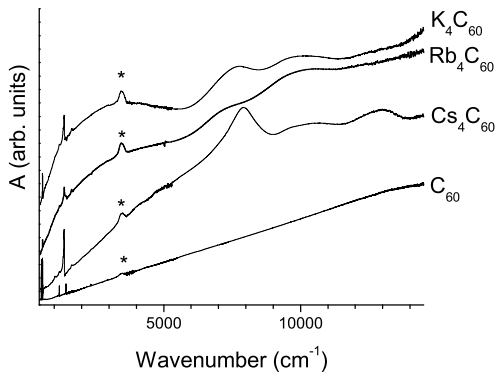}
\caption{Combined MIR-NIR spectra of \K4C60, \Rb4C60 and \Cs4C60
without background correction. The intermolecular transition at
around 4000 \cm-1 (broad background) gets weaker with increasing cation size. (The sharp peaks around 3700 cm$^{-1}$, denoted by asterisks, arise from atmospheric water absorption.)}
\label{fig:mrg}
\end{figure}

\subsection{Electronic transitions}

MIR spectra indicated a molecular symmetry change from \D2h to
\Dhd/\Dod\ on heating in all of the three compounds and electronic
transitions should exhibit similar splitting. A splitting has
been reported in NIR spectra of \K4C60\cite{iwasa95} and was
systematically investigated by transmission electron energy loss
spectroscopy in a series of \A4C60\
compounds.\cite{knupfer96,knupfer97} The KBr pellet technique is
not a particularly good method for quantitative evaluation in a
broad frequency range, due to scattering effects in the pellets
and  inadequate determination of the optical path length. We
nevertheless measured the NIR spectra of all the compounds at
several temperatures and relate our findings to the EELS
measurements by Knupfer and Fink.\cite{knupfer97} Oscillator
strengths in thin film transmission EELS studies can be compared
more reliably between different materials; the frequency
resolution of this method, on the other hand, is only 928 \cm-1\
compared to 4 \cm-1\ in the IR spectra. Therefore, we concentrate
on the number and position of electronic excitations and will not
attempt to draw any conclusion regarding line width or intensity.

Figure \ref{fig:mrg} shows overall (MIR/NIR) spectra of the three
salts and C$_{60}$. It is apparent that there is a finite spectral
weight even at low frequency, and its relative intensity decreases
with increasing cation size. Knupfer and Fink\cite{knupfer97} have
identified this low-frequency excitation around 4000 cm$^{-1}$ ($\sim$ 0.5 eV), as a
transition between Jahn-Teller split states (\emph{e.g.} $e_u \rightarrow
a_{2u}$ in Figure \ref{fig:JTstate}) on \emph{different}
molecules. Intramolecular excitations between JT states are dipole
forbidden, but in a Mott-Jahn-Teller picture, this energy,
renormalized due to intermolecular interactions, becomes the
effective Hubbard repulsion term $U_{eff}$.\cite{fabrizio97} Such
transitions have been observed in one-dimensional organic
conductors.\cite{torrance75}

\begin{figure*}
\includegraphics[scale=0.26,angle=0,trim=70 250 110 -250]{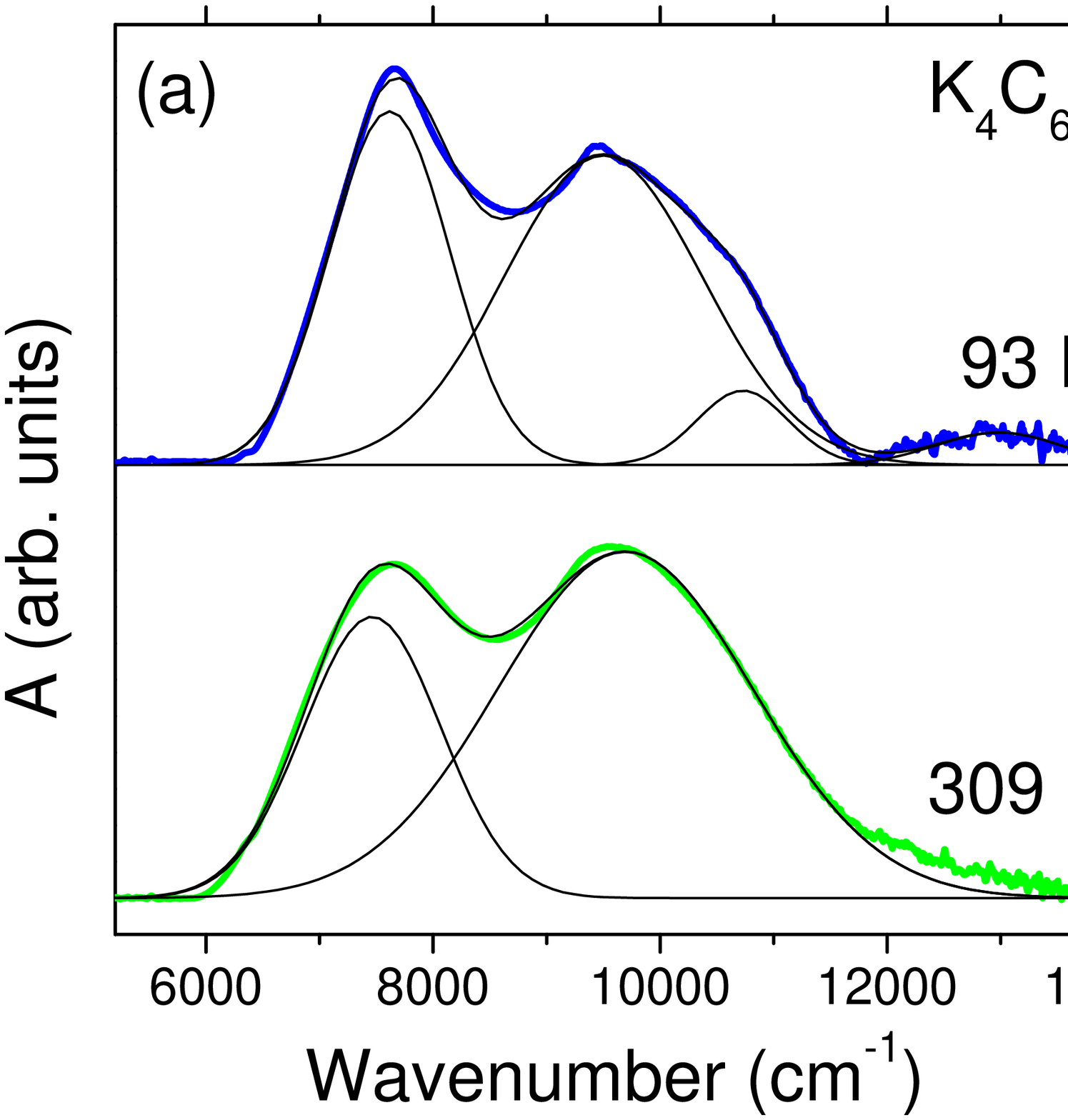}
\includegraphics[scale=0.26,angle=0,trim=70 250 110 -250]{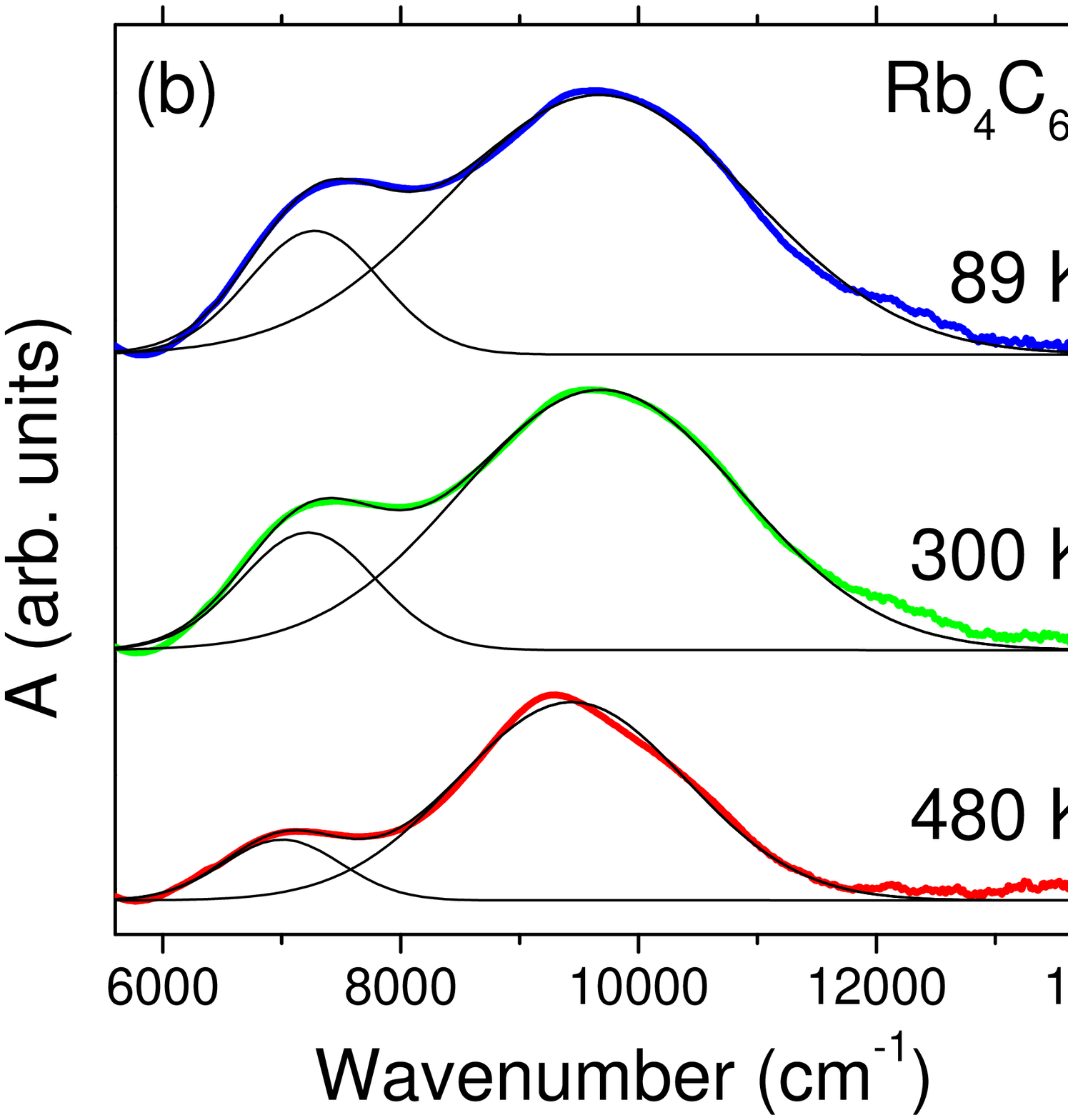}
\includegraphics[scale=0.26,angle=0,trim=70 250 110 -250]{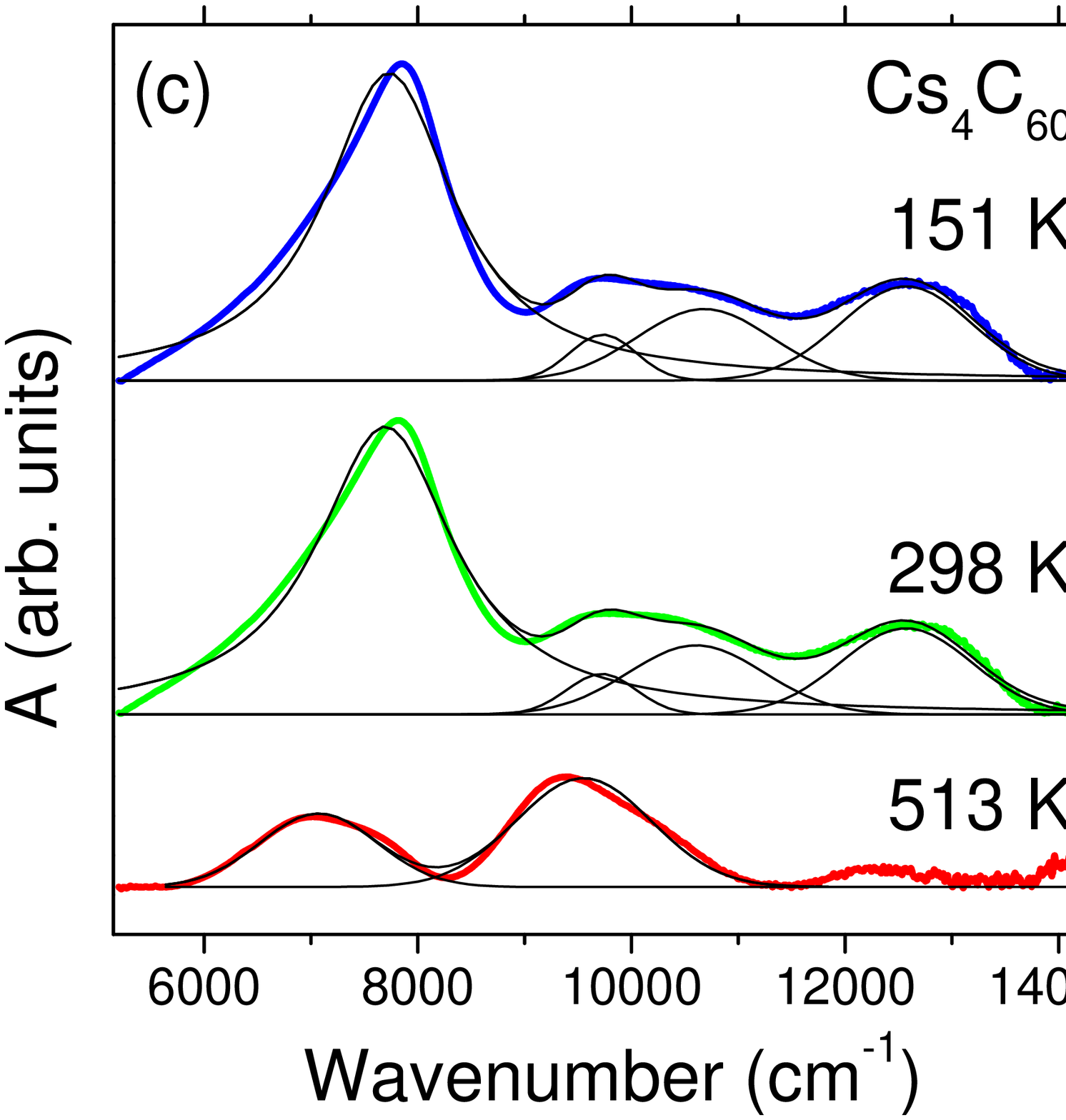}
\caption{(color online) Baseline-corrected NIR spectrum of (a)
\K4C60, (b) \Rb4C60, and (c) \Cs4C60\ at selected temperatures
(colored lines). The spectra were fitted with Gaussians, with the
exeception of the lowest frequency peak of \Cs4C60 at 151~K and
298~K where a Lorentzian produced a better fit. These fits are
shown with black lines.} \label{fig:nir}
\end{figure*}

\begin{table*}
\caption{Peak positions (in \cm-1) of the Gaussians fitted to the
NIR spectra. Note that the lowest frequency peak of \Cs4C60 at
151~K and 298~K is a Lorentzian, which produced a better fit.
\label{tab:nirfit}}
\begin{ruledtabular}
\begin{tabular}{lll|llll|llll}
\multicolumn{3}{c|}{\K4C60} & \multicolumn{4}{c|}{\Rb4C60} &
\multicolumn{4}{c}{\Cs4C60}
\\ 93~K& 309~K & Ref.\onlinecite{knupfer97} & 89~K& 300~K& 480~K &
Ref.\onlinecite{knupfer97} & 151~K & 298~K & 513~K
&Ref.\onlinecite{knupfer97}
\\ \hline 7618$\pm$4 & 7464$\pm$3 & 7421& 7275$\pm$7 & 7224$\pm$3 &
7006$\pm$6& 7904 & 7730$\pm$3 & 7695$\pm$3 & 7071$\pm$6 & 7904
\\ 9499$\pm$10 & 9692$\pm$3 & 10082 & 9671$\pm$7 & 9682$\pm$3 &
9439$\pm$3& 10324 & 9733$\pm$22 & 9719$\pm$24 & 9552$\pm$4 & 10324
\\ 10727$\pm$8 & & & & & & &10680$\pm$50 & 10606$\pm$67 & &
\\ 12982$\pm$15 & & 12582 & & & & 12824 &12571$\pm$19 & 12570$\pm$17 & & 12824
\\
\end{tabular}
\end{ruledtabular}
\end{table*}

In order to better resolve the split NIR lines, we performed a
baseline correction between 6000 and 14000 \cm-1\ and fitted the
remaining lines with Gaussians. The resulting fits are depicted in
Figure \ref{fig:nir}, and the parameters summarized in Table
\ref{tab:nirfit}. Four dipole allowed intramolecular transitions
are expected in the case of \D2h, and two in the case of
\Dhd/\Dod\ (see Figure \ref{fig:JTstate}). This is indeed seen in
\K4C60\ and in \Cs4C60\ and corresponds to the MIR
measurements at all temperatures. However, in
\Rb4C60, while the low-temperature spectra can be fitted with four
Gaussians, the
decomposition was not unambigous since these lines are broad and
their splitting seems to be small. Comparing our parameters with
those reported in Ref. \onlinecite{knupfer97} (Table
\ref{tab:nirfit}), we have the best agreement for \Cs4C60, but
instead of their three peaks we can identify four, as expected
from symmetry. In the case of \K4C60\ Ref. \onlinecite{knupfer97}
found three similar lines as in \Cs4C60, but we see
two at low temperature and four at high temperature. We assume that the discrepancy
originates in the baseline correction of the EELS data for the
higher-lying electronic transition of \cn. (Visual inspection of
the spectra shown in Ref. \onlinecite{knupfer97} reveals that the
1.5 eV peak is much less pronounced in \K4C60 than in \Cs4C60\ and
\Rb4C60.)

To summarize the above, vibrational and electronic spectra in all
three salts indicate $D_{2h}$ distorted C$_{60}^{4-}$ ions at low
temperature and \Dhd/\Dod\ distorted ones at higher temperature at
the time scale of the optical measurements. These methods cannot
distinguish between individual configurations in the static or the
staggered static Jahn-Teller state, nor can they detect
transitions between them. These transitions occur via librational
motion, which can be studied by inelastic neutron scattering.

\subsection{Librations}

Inelastic neutron scattering (INS) spectra were measured as a
function of  momentum transfer, $Q$, and energy transfer, $E$, at
several temperatures for  \K4C60 and \Rb4C60.
Figure~\ref{fig:librations}a and b show spectra at 100~K, 200~K
and 300~K for \K4C60\ and at 100~K and 300~K for \Rb4C60\ at a
constant momentum transfer of $Q=$5.5~\AA$^{-1}$. The solid
symbols are the corrected experimental data and the lines are fits
as described in the experimental details section. Well-defined
peaks are observed at non-zero energy transfer at all temperatures
in both fullerides and may be assigned to librational modes of
\cn\ ions based on the momentum transfer dependence of their
intensities and peak widths. The $Q$-dependence of the  integrated
intensity of the librational modes in fullerides is characteristic
of the form factor of the \C60\ molecule and has been studied in
detail in many fullerides; it provides unambiguous evidence for
the assignment as
librations.\cite{neumann92,christides93,reznik94}
Figure~\ref{fig:librations}c shows $Q$-dependent data at fixed
energy transfers of 2.4~meV and 5~meV for \K4C60. The momentum
transfer spectrum at $E=2.4$~meV is a reasonable substitute for
the $Q$-dependent integrated intensity because the librational peak
position and width are insensitive to $Q$ according to our energy
transfer spectra at a few other selected momentum transfers. The
2.4~meV peak displays the characteristic $Q$-dependence of the
librational modes of fullerides with a small peak around
$Q=$3.5~\AA$^{-1}$ and a larger peak around $Q=$5.7~\AA$^{-1}$.
These peaks are attributed to the non-zero Legendre polynomials
with coefficients of $l=10$ and $l=18$, respectively.\cite{copley92}
In contrast, the momentum transfer spectrum at 5~meV energy
transfer does not show this behavior since it has much less
librational character.

\begin{figure*}
\includegraphics[scale=0.19,trim=-50 0 0 0]{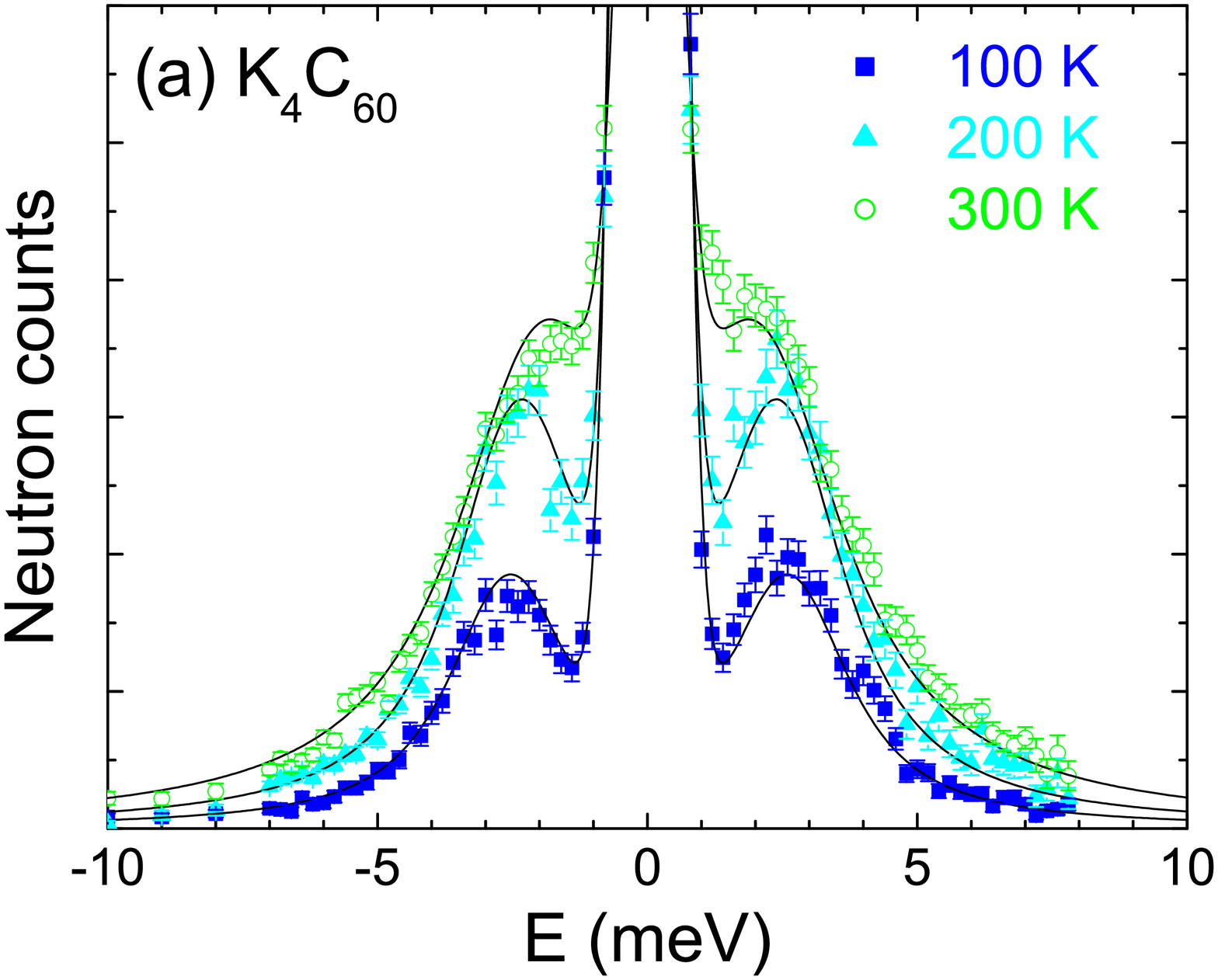}
\includegraphics[scale=0.19,trim=-50 0 0 0]{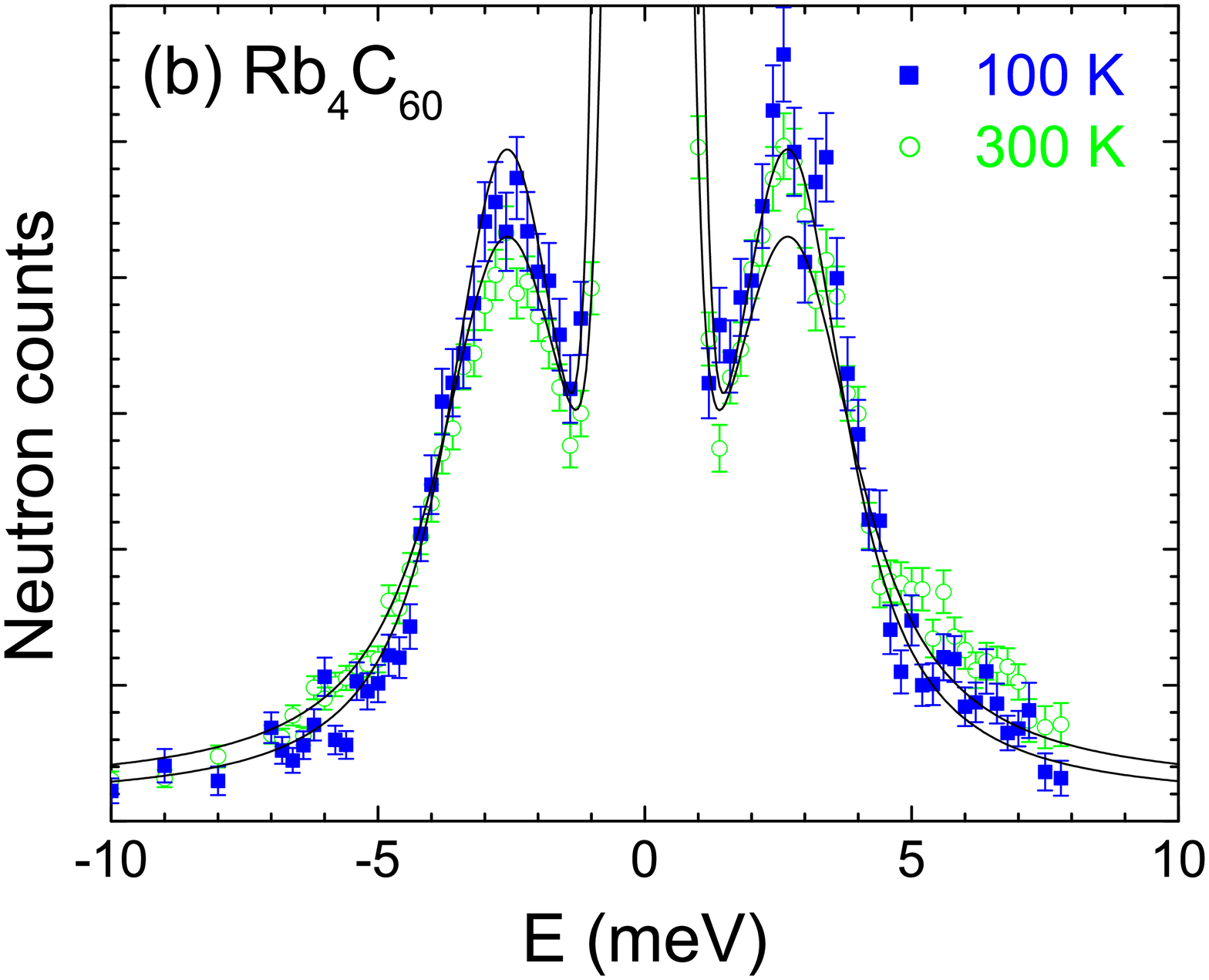}
\includegraphics[scale=0.19,trim=-50 0 0 0]{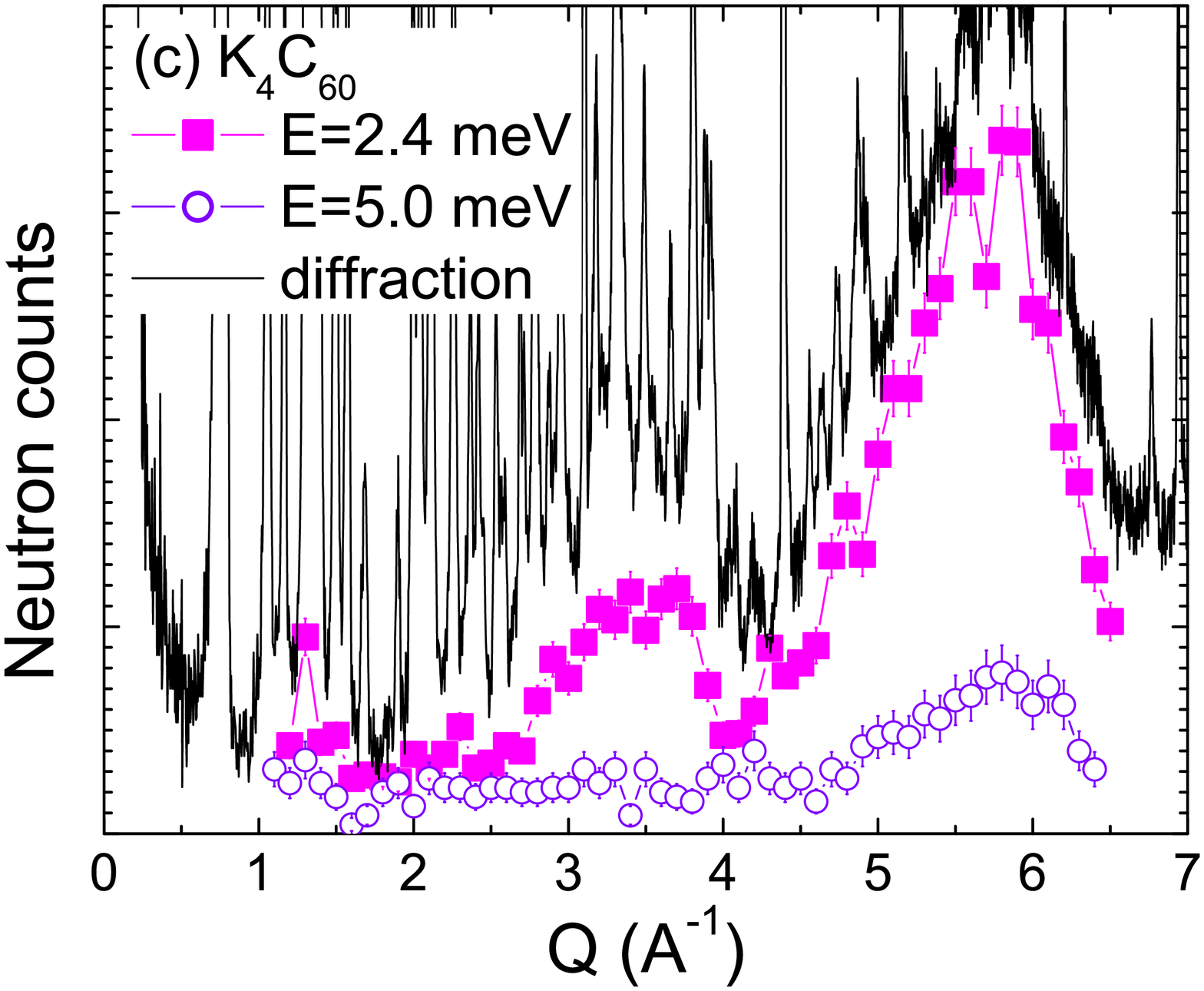}
\caption{(color online) (a) Fixed momentum transfer scans of
\K4C60\ at 100, 200 and 300~K at $Q=$5.5~{\AA}$^{-1}$. (b) Fixed
momentum transfer scans of \Rb4C60\ at 100 and 300~K at
$Q=5.5$~{\AA}$^{-1}$. Symbols are the measured data, lines are
fits as described in the text. (c) Fixed energy transfer scans at
$E=$2.4~meV and $E=$5~meV of \K4C60\ along with the neutron
diffraction pattern scaled to emphasize the diffuse background.}
\label{fig:librations}
\end{figure*}

The main motivation for the low temperature diffraction
experiments was to search for a possible structural phase
transition similar to the order-disorder transition in
\Cs4C60.\cite{dahlke02} Additional proof that there is no ordering
of the \cn\ between room temperature and 6~K in \K4C60\ is
presented by the large and temperature independent diffuse
background. Figure~\ref{fig:librations}c compares the diffuse
background to the fixed energy transfer scans discussed above. As
observed for other fullerides, the $Q$-dependence of the diffuse
background of the diffraction is very similar to that of the
librational peak, indicating disorder of the \cn\
anions.\cite{christides92}

\begin{table}
\caption{Measured librational energies ($E_{lib}$) of \K4C60 and
\Rb4C60, calculated energy barriers of the
reorientation ($E_{barrier}$) and root mean square
librational amplitudes ($\Theta_{rms}$) at 100~K and 300~K.}
\begin{ruledtabular}
\begin{tabular}{l|ccc}
 &$E_{lib}$ (meV) & $E_{barrier}$  (meV) & $\Theta_{rms}$  (\deg)
\\ \hline \K4C60\ 100~K &2.57 $\pm$ 0.10 &277$\pm$22 & 3.55 $\pm$ 0.01
\\ \K4C60\ 300~K &2.00 $\pm$ 0.10 &168$\pm$17 & 7.85 $\pm$ 0.02
\\ \hline \Rb4C60\ 100~K &2.65 $\pm$ 0.10 &294$\pm$22 & 3.44 $\pm$ 0.07
\\ \Rb4C60\ 300~K &2.64 $\pm$ 0.10 &293$\pm$22 & 5.96 $\pm$ 0.01
\\
\end{tabular}
\label{table:nisdata}
\end{ruledtabular}
\end{table}

The librational energies obtained by fitting the inelastic peaks
can be found in Table~\ref{table:nisdata}. Following the arguments
of Neumann \emph{et al.},\cite{neumann92} the rotational barrier
between the two orientations can be estimated assuming that a
simple sinusoidal hindrance potential can describe the rotational
motion of the \C60 anion. For small amplitudes of libration, the
potential barrier is calculated as
\[
E_{barrier}=\frac{E_{lib}^2}{B}\left(\frac{\Theta_{jump}}{2\pi}\right)^2
\]
where $\Theta_{jump}$ is the reorientation angle between
neighboring potential minima, $B=0.364~\mu$eV is the rotational
constant for \C60, and $E_{lib}$ is the librational energy at a
given $Q$ and temperature. For $\Theta_{jump}$ we assume
44.5$^{\circ}$, meaning that a rotation about the $C_3$ axis of
the molecule -- which is approximately in the [111] direction --
brings the molecule from one standard orientation to the other.
We obtain a
value for the potential barrier $E_{barrier}=277$~meV for \K4C60
and 294~meV for \Rb4C60 based on the observed $E_{lib}$ at 100~K.
These estimated potential barriers are comparable to \C60\ and
much smaller than in K$_3$\C60\ or Rb$_6$\C60,\cite{neumann93}
indicating smaller crystal fields in \A4C60. The smaller crystal
field is a consequence of the larger free volume in the \A4C60
compounds compared to  A$_3$\C60\ or A$_6$\C60.\cite{sd04}

The mean amplitude of the libration
can also be calculated within the harmonic approximation from the librational energy via

\[
\Theta_{rms}= \sqrt{\frac{4B}{E_{lib}}\coth\left(\frac{E_{lib}}{2kT}\right)}
\]

The obtained $\Theta_{rms}$ values are shown in Table
\ref{table:nisdata}. The $\Theta_{rms} = 7.8^{\circ}$ value of
\K4C60 at room temperature is fairly large, which is readily seen
when comparing it with the 7$^{\circ}$ value of \C60 near its
phase transition.\cite{copley92} For \C60, 7$^{\circ}$ is considered the
critical angle for orientational melting. There are
additional similarities between the temperature dependence of the
librational peak in \K4C60 and \C60. The librational mode of
\K4C60 softens and widens with increasing temperature as in \C60
\cite{copley92} and in the monomer phase of Na$_2$Rb\C60  below
their phase transition temperature during which the rotation of
 the molecules becomes free.\cite{copley92,tanigaki94,christides92} In \Rb4C60 at room
temperature, the librations increase only to 6$^{\circ}$ and the
other trends are also absent.

Based on the above similarities in the temperature dependence of
the librations between \K4C60 and \C60 we raise the possibility
that \K4C60 is close to an orientational melting transition at
room temperature. This transition would be in accordance with the
observed change of symmetry in the motion of \cn\ found at 250~K
by NMR in \K4C60.\cite{Zimmer94} High temperature INS experiments
are planned in order to search for this transition.

\section{Discussion}

Two separate effects determine the distortion of fulleride anions
in a lattice: the JT effect of the molecule and the  crystal field
of the external potential caused by the counterions. Our
structural and spectroscopic results help to determine the
relative importance of these two effects depending on cation size
and temperature. We also discuss the importance of the two dynamic
processes, pseudorotation and molecular reorientation, based on
spectroscopy and neutron scattering.

The low-temperature phase of all three \A4C60\ salts studied can
be modeled by the constructive interaction of the JT effect and
the external potential, resulting in $D_{2h}$ distorted fulleride
ions. The $D_{2h}$ molecular point group is identical to the crystal space
group $Immm$ ($D_{2h}^{25}$) of \Cs4C60\ and the largest common
subgroup of $I_h$ and $I4/mmm$ ($D_{4h}^{17}$) of \K4C60\ and
\Rb4C60. Accordingly, \Cs4C60\ forms a true cooperative static
Jahn-Teller state and the other two salts a static Jahn-Teller
state with distorted ions randomly occupying the two standard
orientations. Since the molecular symmetry is identical and the
molecules are static at the time scale of the spectroscopic
measurements, vibrational and electronic spectra are independent
of the crystal structure. The reason why these structures are
different has been given by Dahlke \emph{et al.}\cite{dahlke02}
following Yildirim \emph{et al.}:\cite{yildirim93b} to minimize
repulsive interaction between cations and anions due to orbital
overlap.\cite{fischer93} According to this model, the
orientational order in the orthorhombic phase of \Cs4C60 appears
to avoid close Cs--C contacts, which would arise in the disordered
structure.\cite{dahlke02} In the other two compounds where the
free volume is larger, the two standard orientations remain but
reorientation between them slows down. Dahlke \textit{et
al.}\cite{dahlke02} estimated the critical value of the
controlling parameter (closest cation-anion center distance minus the
cation radius) to fall between the low-temperature phase
of \Cs4C60\ and \Rb4C60. With increasing temperature,
\Cs4C60\ reaches this critical value and a phase transition to a
tetragonal phase happens between 300 K and 623 K.\cite{dahlke02}
The crystal structure of
\Cs4C60\ at high temperature and \K4C60\ and \Rb4C60\ at all
temperatures are similar.\cite{dahlke02, kuntscher97} According to our infrared results, the
\emph{molecular} point group in each compound is changing from
\D2h\ to \Dhd/\Dod\ upon warming, the transition temperature
increasing with cation size. Lacking structural data at
intermediate temperatures, we cannot tell whether the symmetry
change in \Cs4C60\ coincides with the structural transition, but
in \K4C60\ and \Rb4C60, we definitely observe a change of
molecular geometry \emph{without} changing the crystal structure.

The \D2h distortion can only be realized when an external
potential, like that of the surrounding cations, lowers its
energy. As the temperature is raised, the lattice expands, and at
the same time pseudorotation becomes more probable, both
contributing to a competition between the molecular Jahn-Teller
effect and the external potential. As the molecular ions decouple
from the lattice, the tendency to behave as isolated ions gets
stronger and thus the possibility of \Dhd/\Dod\ distortions
increases. The estimated distortion of \cn\ ions is the largest
among the fulleride ions, larger than in
C$_{60}^{3-}$,\cite{dahlke02} which further explains the
difference in electronic properties between the two types of
compounds. The significance of this effect relative to the crystal
field increases with increasing temperature and increasing cation
-- anion distance. The scaling of the transition
temperatures with cation size corroborates this assumption.

In the following we consider four possible structural models,
depicted schematically in Figure \ref{fig:modell}. The dark
background symbolizes the volume into which the molecule is
confined by the crystal field; the growth of this area from model
(1) to (4) indicates a decreasing strength of the crystal field
due to heating or smaller cation size. The light blue areas
represent the fulleride ions; the direction of the minor axis of
an ellipse refers to the direction of the principal molecular
axis. In the case of a \D2h distortion this principal axis
intercepts two hexagon-hexagon bonds, while in the case of \Dhd\
and \Dod\ distortions it goes through the centers of two hexagons
and two pentagons, respectively (see Figure \ref{fig:distort}a).
Thus the direction of the principal axis determines the point
group of the molecule standing in a given orientation. This way
the horizontal ellipse in the figure of model (3) corresponds to
the pancake-shaped \D2h distortion. Ellipses in other directions
should be considered as representing \Dhd\ or \Dod\ distortions.

\begin{figure}

\begin{center}
\includegraphics[scale=0.38,angle=-90,trim=45 80 0 0]{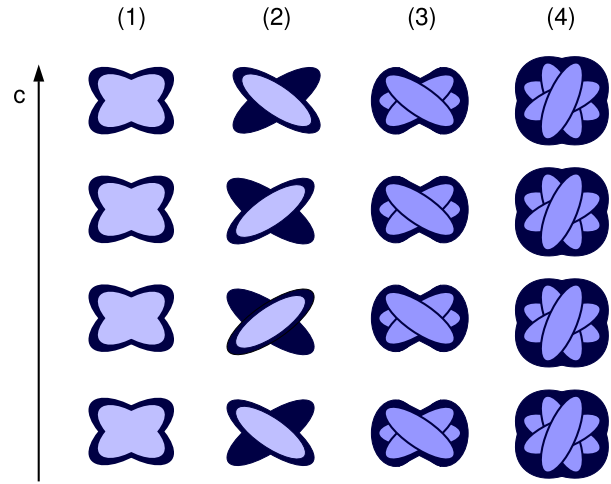}
\caption{(color online) Possible structural models of Jahn--Teller
distorted fulleride ions (light blue) in a crystal field. The dark
background symbolizes the volume in which the molecule is confined
by the crystal field. The figures within each column depict
fulleride ions on different lattice points. The magnitude of the
distortion of the fulleride ions is overemphasized for clarity.
(1) Static \D2h\ distortion, following the shape of the crystal
field. (2) Static \Dhd/\Dod\ distortion. The direction of the
distortions can be ordered (staggered static Jahn-Teller state) or
disordered. (3) Dynamic distortion in preferred directions
appearing in an anisotropic crystal field. Both pancake-shaped
\D2h\ and \Dhd/\Dod\ distortions are realized with time. (4) Free
pseudorotation of the \Dhd/\Dod\ distortion in a weak isotropic
crystal field. The deeper blue of the molecules in models (3) and
(4) represents a temporal average. This picture illustrates only
the distortion of the anions; for their orientation see Figure
\ref{fig:JTstate}.} \label{fig:modell}
\end{center}
\end{figure}

If the crystal field is very strong (model (1)), it causes a
static \D2h\ distortion. It is possible, although not probable,
that this distortion is identical to the pancake shape
corresponding to the saddle point on the APES of isolated
molecules. In a general case -- like that of \Cs4C60 -- the
distorted molecule has a different shape. The clover shape of the
light blue region in the figure represents that of the anion found
in \Cs4C60 by neutron diffraction.\cite{dahlke02} Model (1) works
for both orthorhombic and tetragonal crystal field since the
molecular point group is the same in both cases. The tetragonal
crystal structure requires the average of the atomic positions
over the crystal to exhibit D$_{4h}$ symmetry. As it has been
mentioned earlier, one molecule cannot distort into this point
group, but the spatial average of \D2h\  molecules randomly
distributed over two standard orientations will produce the
required fourfold axis.

In models (2) and (3), the crystal field is weaker than in the
previous case, thus it allows the appearance of the squeezed
\Dhd/\Dod\ distortions, which are the minima of the molecules'
APES. The conversion between equivalent \Dhd/\Dod\ distortions
with the principal axis pointing in different directions could
take place by pseudorotation through a pancake-shaped \D2h\
distortion.

In model (2), the intermediate \D2h\ distortion still leads to too
short A--C distances, thus pseudorotation is not possible and the
distortion is static. The distorted molecules can be arranged in
the crystal either ordered in some way (staggered static state) or
totally disordered regarding the direction of their principal
axis.

In model (3) the crystal field is considerably weaker in some
directions (e.g. in $a$ and $b$) than in others, thus molecules
can extend in these directions. This way distortions can appear
not only in different directions but also with different point
groups. The pancake-shaped \D2h\ distortion is present as the most
favored distortion of the crystal field and the \Dhd/\Dod\
distortions are present because they are preferred by the
molecular JT effect. Although the molecule is not free to take up
distortions in any direction, the allowed distortions can convert
dynamically among themselves. The MIR spectrum of this state would
contain five lines originating from each \T1u molecular mode:
three corresponding to \D2h\ molecules and two to \Dhd/\Dod\
molecules. Similarly, sixfold splitting should appear in the NIR
spectrum. Since fitting spectra with many more parameters
invariably yields a better fit, we cannot distinguish between
states with only \D2h\ and with both \D2h\ and \Dhd/\Dod\
distortions.

In model (4) the crystal field is very weak, thus the molecule can
perform free pseudorotation in the crystal just like an isolated
molecule. As the potential is very nearly isotropic, the
pancake-shaped \D2h\ distortion is no longer favored, and only the
\Dhd/\Dod\ distortions appear.

Models (3) and (4) contain dynamical disorder of
distorted molecules. IR spectroscopy only detects the individual
distortions and not their average if the timescale of the
spectroscopic excitations is smaller than the time scale of
pseudorotation.

The low-temperature phase of the three \A4C60\ salts corresponds
to model (1), containing statically \D2h\ distorted molecules due
to the strong crystal field. In \Cs4C60 the abrupt change of the
crystal field at the phase transition can result in a simultaneous
change of the molecular distortion, to any of the models (2), (3)
or (4). Further heating will lead to states with gradually
weakening crystal field, in the order: model (2) $\longrightarrow$
model (3) $\longrightarrow$ model (4). In \K4C60 and \Rb4C60 the
absence of a phase transition indicates a continuous transition
from model (1) to models with \Dhd/\Dod\ molecules. Such a
continuous transition cannot lead from model (1) to model (2),
though. The explanation is as follows. The possible configurations
of a molecule in a crystal are those of the isolated molecule
(corresponding to the lowest energy points of the warped trough of
the APES), and those preferred by the crystal field (\D2h
distortions with the shape preferred by the surroundings of the
molecule). A continuous transition can lead from the former to the
latter only if there is no high energy barrier between them. The
intermediate configurations are the \D2h\ saddle points on the
trough of the APES. As these configurations have high energy in
model (2), no continuous transition can lead to this state.
Because model (3) and (4) contain dynamically distorted molecules,
we conclude that on heating a static-to-dynamic transition takes
place in \K4C60 and \Rb4C60.

Pseudorotation is not to be confused with molecular reorientation
which we studied by inelastic neutron scattering. From the
molecular point of view this motion is an abrupt rotation of the
crystal field. During the rotation the distortion of the molecule
should follow the change of the crystal field. Thus in the two
standard orientations the direction of the distortions is
different. INS data complement the spectroscopic results in two
ways: they emphasize the possibility of the rotational motion
around a $C_3$ axis, thus stressing its importance, and they prove
the weakening of the crystal field with increasing temperature,
through increasing librational amplitudes.

The results shown here are in good agreement with the $^{13}$C-NMR spectra of Ref. \onlinecite{brouet02}. Above 150 K the reorientational motion observed in our NIS measurement could correspond to a rotation around one of the four $C_3$ axes of the molecule on the long time scale of NMR measurements. Thus when the axis of this rotation changes with a lower frequency than that of the NMR measurement, it causes the observed NMR lineshape characteristic of uniaxial motion.\cite{brouet02} Below 150 K the reorientational motion could be static on the NMR time scale leading to the observed line broadening. Around 250 K the shape of the NMR line changes, which could correspond to the changing of the molecular symmetry from model (1) to model (3) or (4).

\section{Conclusions}

MIR and NIR measurements showed that the same molecular geometry
change is present in \K4C60, \Rb4C60 and \Cs4C60: the point group
of the \cn\ molecular ion changes from \D2h to \Dod\ or \Dhd\ on
heating. Contrary to \Cs4C60, where an orthorhombic-tetragonal
transition takes place, we did not find a structural phase
transition in \K4C60 and \Rb4C60. The absence of a phase
transition can be explained by the smaller cation-anion overlap
which does not stabilize the orthorhombic structure.

Since the molecular geometry change in \K4C60 and \Rb4C60 is not
coupled to a phase transition, the fundamental role of the
molecular Jahn--Teller effect in the transition is obvious. On
heating, the importance of the Jahn--Teller effect is increasing
as the \D2h potential of the surrounding cations decreases and the
number of accessible degrees of freedom increases. The weakening
of the crystal field on heating is also indicated by the INS
results.

Because of the dominance of the crystal potential in the \D2h
distortion, this distortion is static. In the case of  \K4C60
and \Rb4C60 we suggest that a dynamic Jahn-Teller state develops
as the \Dhd/\Dod\ distortions appear.

From the splitting of the electronic transition around 1~eV we
conclude that the energy bands in the solid reflect the
Jahn-Teller distortion of the molecular ions; the presence of the
0.5~eV feature, which is forbidden in the molecule and therefore
must be assigned to intermolecular excitations, signals the
importance of electron-electron correlations in the solid. We
regard the simultaneous appearance of these two features as
experimental proof of the Mott-Jahn-Teller insulator
state.\cite{fabrizio97}

\begin{acknowledgments}

We thank G\'abor Oszl\'anyi for his invaluable help with characterizing the samples by X-ray diffraction and
most enlightening conversations. We also gratefully acknowledge useful discussions with Dan Neumann and Terrence Udovic. Financial support was provided by OTKA grants T 034198 and T 049338 and NSF-INT grant 9902050.

\end{acknowledgments}

\bibliography{cikk}

\end{document}